\documentclass[11pt,a4paper]{article}
\pdfoutput=1

\usepackage[T1]{fontenc}
\usepackage[utf8]{inputenc}

\usepackage{jcappub}

\usepackage{mathtools}         
\usepackage{bm}                
\usepackage{booktabs}          
\usepackage{multirow}          
\usepackage{tabularx}          
\usepackage{siunitx}           
\usepackage{subcaption}        
\usepackage{xspace}            
\usepackage{float}
\usepackage{placeins}
\usepackage{graphicx}
\usepackage{placeins}

\title{Impact of Primordial Magnetic Field-induced Small-scale Power on 
Non-linear Structure Formation and Cosmic Reionization}

\author[a]{Hiten,}
\author[b]{Kanhaiya Lal Pandey,}
\author[a]{Rajesh Mondal,}
\author[c]{Subinoy Das,}
\author[c]{Sharanya Sur}

\affiliation[a]{Department of Physics, National Institute of Technology,
Calicut, Kerala -- 673601, India}

\affiliation[b]{Department of Physics, School of Advanced Sciences, 
Vellore Institute of Technology, Vellore, 632014, Tamil Nadu, India}

\affiliation[c]{Indian Institute of Astrophysics,
Bengaluru, Karnataka -- 560034, India}

\emailAdd{hiten.bhankar05@gmail.com}
\emailAdd{kanhaiyalal.pandey@vit.ac.in}
\emailAdd{rajeshmondal@nitc.ac.in}
\emailAdd{subinoy@iiap.res.in}
\emailAdd{sharanya.sur@iiap.res.in}

\abstract{Primordial magnetic fields (PMFs) induce additional small-scale matter density perturbations that can significantly alter halo formation and cosmic reionization. We investigate this structure formation using a PMF-modified initial matter power spectrum in collisionless $N$-body simulations. The resultant halo catalogues supply the ionizing sources for a semi-numerical reionization calculation, enabling the construction of ionization and 21-cm brightness-temperature fields in the saturated spin temperature limit. Our main analysis compares the $\Lambda$CDM reference model against four PMF models with magnetic field amplitudes $B_0=100$, $130$, $150$, and $170\,\mathrm{pG}$ at a fixed magnetic spectral index of $n_B=-2.9$. Holding the background cosmology and source prescription fixed, we find that the PMF models retain enhanced small-scale matter clustering, increase the abundance of early low-mass haloes, and accelerate reionization. The volume-averaged reionization midpoints shift by $\Delta z_{50} \simeq 0.50$ to $2.14$ relative to $\Lambda$CDM, accompanied by changes in ionization morphology and 21-cm power spectra. Consequently, the Thomson optical depth increases from $\tau_{\rm e} \simeq 0.0547$ in the reference model to $0.0595$--$0.0786$ across the PMF models. These differences correspond to $0.66$--$3.28$ times the Planck 2018 uncertainty ($\sigma_\tau=0.0073$), which represents simulation relative offsets rather than formal exclusion limits. The source efficiency $N_{\rm ion}$ is held fixed across models and calibrated to achieve complete reionization near $z\sim6$ and CMB Thomson optical depth close to Planck 2018 in the referenced model, allowing us to examine the response to PMF-modified initial clustering without re-tuning the source prescription. Our calculations do not follow the subsequent evolution of magnetic fields, nor do we include dissipative magnetic heating. Translating these sensitivities into observational bounds will require treatment of astrophysical degeneracies and numerical uncertainties.}


\begin{document}
\maketitle
\raggedbottom





\section{Introduction}
The Epoch of Reionization (EoR) offers a sensitive probe of the small-scale structure of the standard cosmological model. During this epoch, ultraviolet radiation from early galaxies ionized the intergalactic medium (IGM) from mostly neutral to fully ionized state. The timing and spatial development of this transition depend on the abundance and clustering of source haloes, as well as on the efficiency with which they produce and release ionizing photons. In this way, reionization connects primordial fluctuations and nonlinear gravitational collapse to observable properties of the high-redshift Universe. The redshifted 21-cm line of neutral hydrogen provides a particularly rich probe of this connection, encoding detailed information about the density, thermal state and ionization structure of the IGM~\cite{Furlanetto2006, Pritchard2012}. Consequently, departures from the standard small-scale matter power spectrum can leave detectable imprints on reionization even when their effects on large-scale clustering are negligible\cite{Sarkar2016, Das2018}.

Primordial magnetic fields (PMFs) are a well-motivated source of such departures. Fields generated during inflation or cosmological phase transitions would retain information about high-energy physics and could serve as seeds for the magnetic fields observed in astrophysical systems~\cite{Durrer2013, Subramanian2016}. Magnetic forces can generate additional matter density perturbations \cite {Adi2023PMFDensity}, while magnetic energy dissipation through ambipolar diffusion and decaying magnetohydrodynamic turbulence can modify the thermal and ionization histories of the IGM \cite{Sethi2005, Bhaumik2025}. Cosmic Microwave Background (CMB) measurements constrain stochastic PMFs, although inferred limits depend on the assumed spectral shape, helicity, and physical effects included~\cite{PlanckPMF2015}. High-redshift structure formation and reionization offer complementary constraints because they respond directly to small-scale fluctuations and to the formation of the earliest ionizing sources.

After recombination, Lorentz forces source the baryonic density perturbations whose gravitational potentials subsequently drive fluctuations in the dark matter field~\cite{Sethi2005, Adi2023PMFDensity}. For statistically uncorrelated adiabatic and magnetically sourced modes, these contributions add to the linear matter power spectrum. The resulting excess in small-scale power increases the abundance of low-mass haloes and the total collapsed mass fraction that is available for star formation. The amplitude and shape depend jointly on the comoving field strength $B_0$, the magnetic spectral index $n_B$, and the treatment of radiative damping and magnetic-pressure suppression. Therefore, we can say that the field amplitude alone does not determine the response of the ionizing-source population. Magnetic heating through ambipolar diffusion and decaying turbulence provides a separate channel by altering the gas temperature and its ability to accrete into small host haloes~\cite{Sethi2005, Subramanian2016, Bhaumik2025}. Distinguishing between the structure formation and thermal channels is essential when translating reionization observables into constraints on PMF parameters.

Connections between PMFs and reionization have been studied using analytic and semi-numerical approaches. The effect of PMF on the 21-cm fluctuations has been previously studied \cite{TashiroSugiyama2006_21cm} including the signal from the time of reionization \cite{SethiSubramanian2009PMFHI}. Subsequently, Pandey et al.~\cite{Pandey2015} combined semi-numerical modeling of reionization and observational data to jointly constrain magnetic-field and astrophysical parameters. More recently, Adi, Cruz and Kamionkowski~\cite{Adi2023PMFDensity} revisited the derivation of the magnetically sourced density power spectrum, solving important discrepancies in previous calculations. Using this formalism, Cruz et al.~\cite{Cruz2024} used \texttt{21cmFAST}~\cite{Mesinger2011_21cmFAST} to investigate the influence of PMF on the global 21-cm signal and its power spectrum. We find that an increased early structure formation leads to larger stellar radiation backgrounds, modifies the reionization history and leads to potentially observable signatures, with the CMB optical depth providing an additional consistency check. In this framework, however, the PMF contribution to halo formation is propagated via first-order perturbation theory implemented in \texttt{21cmFAST}, rather than through an explicit $N$-body treatment that directly resolves the non-linear gravitational dynamics of the PMF-modified density field.

This difference is the motivation for our work. We evolve the PMF-modified initial conditions using a particle-mesh $N$-body solver \cite{Mondal2015} and identify haloes from the resulting particle distribution using a Friends-of-Friends (FoF) halo finder \cite{Mondal2016EoRCovariance}, yielding a direct non-linear realization of the magnetically enhanced halo population without recourse to analytic mass-function extrapolations. We also \emph{intentionally restrict} our calculation to the structure-formation channel by excluding magnetic heating via ambipolar diffusion and decaying MHD turbulence. Thus, the induced shifts in CMB Thomson scattering optical depth $\tau_{\rm e}$ and reionization midpoint advance provide a \emph{conservative bound} to the full PMF reionization signal.

The present study has a direct bearing on HERA and SKA observations, in light of the recent work by Worku, Cruz and Kamionkowski~\cite{Worku2026}. They adopted the analytic model \texttt{zeus21} incorporating magnetically sourced small-scale power, radiative damping and magnetic-pressure suppression to probe changes to early galaxy formation and to predict the 21-cm signal, emphasizing correlated changes to Lyman-$\alpha$ coupling, X-ray heating and ionization driven by a modified source population. Like Cruz et al.~\cite{Cruz2024}, their framework propagates the PMF power spectrum through analytic approximations for the collapsed fraction, rather than a direct $N$-body realization of halo formation. In addition, Bhaumik, Paul, \& Pal~\cite{Bhaumik2025} make further predictions that emphasize the necessity of incorporating magnetic heating along with intensified star formation. These studies together motivate the central question we ask here: how does a PMF-modified matter spectrum propagate through \emph{non-linear gravitational evolution} into a directly simulated halo population and reionization field?

This question calls for careful attention to the non-linear evolution and to the numerical resolution. Halo-scale modes collapse gravitationally and the late-time dynamics of this process cannot be deduced from linear perturbation theory. Hydrodynamical simulations by Ralegankar, Pavi\v{c}evi\'c \& Viel~\cite{Ralegankar2024ConsistentIC} examined PMF-consistent initial conditions and the differing responses of baryons and dark matter, while subsequent MHD calculations~\cite{Ralegankar2025nonlinearMPS} showed that magnetic back reaction near the linear-to-nonlinear transition can substantially modify the predicted power spectrum. Lyman-$\alpha$ forest analyses provide further evidence of the sensitivity of high-redshift observables to magnetically induced small-scale power~\cite{Pavicevic2025}.

The motivation for resolving these small-scale features is particularly strong for the upcoming Square Kilometer Array (SKA) observations. Statistical measurements of the 21-cm field will constrain the evolution and characteristic scales of ionized regions, and with sufficiently high-sensitivity imaging one will be able to map their spatial morphology \cite{Iliev2015}. These observables complement the CMB Thomson optical depth $\tau_e$, which measures the line-of-sight (LoS) integrated free-electron column density but does not uniquely determine the redshift dependence or topology of reionization. Changes in halo-scale power alter the source distribution, transmitting the PMF signature into larger ionized structures. Combining the redshift evolution and scale dependence of the 21-cm signal with CMB constraints offers a promising route to testing PMF scenarios~\cite{Cruz2024, Worku2026}, provided that instrumental systematics and astrophysical degeneracies (such as source efficiency, minimum halo mass, heating, and recombinations) are properly accounted for.

In the present work, we have connected the non-linear structure formation response of the PMF-modified initial matter power spectrum to spatially resolved EoR observables. We evolve the initial Gaussian realizations with a collisionless particle-mesh $N$-body solver, identify source haloes using friends-of-friends algorithm, and then pass the resulting halo catalogues to \texttt{ReionYuga} which then constructs the ionization fields and 21-cm brightness-temperature maps~\cite{Mondal2017EoRCovariance}. This approach builds on previous simulation studies linking non-standard small-scale power to reionization~\cite{Sarkar2016, Das2018}. We hold the background cosmology, initial random phases, and ionizing-source efficiency fixed across all runs; changes in the non-linear matter power spectrum, halo mass function, neutral-hydrogen morphology, 21-cm power spectrum, global neutral fraction, and Thomson optical depth can be cleanly attributed to the PMF-modified initial power spectrum. This construction isolates the non-linear gravitational response while leaving dynamical field evolution and magnetic heating outside the calculation. The 21-cm predictions are evaluated in the saturated-spin-temperature limit.

The structure of the paper is as follows. Section~2 describes the theoretical formulation for the PMF-induced matter power spectrum and the magnetic-pressure suppression scale used. In Section~3 we describe the $N$-body simulations, halo identification and semi-numerical \texttt{ReionYuga} reionization modeling. We present the results in Section~4, including the evolved non-linear matter power spectra, the halo mass functions, the 21-cm statistics and the optical-depth comparisons to the $\Lambda$CDM predictions. Finally, Section~5 summarizes and discusses, and proposes the necessary extensions to observational inference.


\section{Primordial Magnetic Field Induced Matter Power Spectrum}
\label{sec:pmf-review}
We assume a stochastic, Gaussian, non-helical primordial magnetic field generated during or before the radiation-dominated era. Its energy density is treated as a first-order perturbation of the homogeneous Friedmann–Lemaître–Robertson–Walker (FLRW) background. Due to the high electrical conductivity of the primordial plasma, the magnetic field is effectively frozen into the plasma on scales where ideal magnetohydrodynamics applies. Consequently, magnetic flux conservation implies that the physical magnetic field scales with the expansion of the Universe, as $\mathbf{B}(\mathbf{x},t)={\mathbf{B}(\mathbf{x})}/{a^2(t)}$, where $a(t)$ is the cosmological scale factor and $\mathbf{B}(\mathbf{x})$ denotes the magnetic field at the present time. 

For a statistically homogeneous and isotropic, non-helical field, the two-point correlation function in Fourier space can be written as,
\begin{equation}
\langle B^\star_i({\mathbf k})B_j({\mathbf k'})\rangle =(2\pi)^3 \delta_D^{(3)} ({\mathbf k}-{\mathbf k'}) P_{ij}({\mathbf{\hat k}}) P_{\rm B}(k)\,, 
\label{spectrum}
\end{equation}
where spatial indices $i,j \in (1,2,3)$, $\hat{k}_i=k_i/k$ is the unit wave vector, $P_{ij}({\mathbf{\hat k}})=\delta_{ij}-\hat{k}_i\hat{k}_j$ acts as the transverse plane projector, $\delta_D^{(3)}({\mathbf k}-{\mathbf k'})$ is the three-dimensional (3D) Dirac delta function. The power spectrum of the magnetic field, $P_B (k)$ is assumed to follow a simple power law, $P_{\rm B}(k) = A_Bk^{n_{\rm B}}$, where $A_B$ normalizes the power in magnetic fields and $n_{\rm B}$ is the spectral index. The amplitude $A_{\rm B}$ is determined by defining the rms value of the smoothed magnetic field over a chosen comoving length scale $\lambda$, $B_\lambda= \langle \mathbf{B}(\mathbf{x})\cdot \mathbf{B}(\mathbf{x})\rangle^{1/2}\vert_{\lambda}$. It can be shown that, $B_\lambda$, obtained using a Gaussian kernel function $W_\lambda(k)=e^{-k^2\lambda^2/2}$, can be expressed as, 
\begin{equation}
B^2_\lambda = \int_0^{\infty}\frac{d^3k}{(2\pi)^3} P_{ii}(\hat{\mathbf k})P_B(k) e^{-k^2\lambda^2} = \frac{2A_B \Gamma\left(\frac{n_B+3}{2}\right)}{(2\pi)^2 \lambda^{n_B+3}} \,. 
\end{equation}
For the smoothing scale $\lambda$ corresponding to $k_\lambda = 1 {\rm Mpc}^{-1}$, we obtain 
\begin{equation}
\label{eq:B0_norm}
A_{\mathrm B} = \frac{(2\pi)^{n_{\rm B}+5} B_0^2}{2\Gamma[(n_{\rm B}+3)/2]}\,,
\end{equation}
where $B_0$ represents the rms strength of the smoothed magnetic field over a scale $k_\lambda = 1 {\rm Mpc}^{-1}$ \cite{2002PhRvD..65l3004M,2007PhRvD..75b3002K,2024A&A...690A..59S,2024ApJ...972..117Z}. The spectral index $n_{\rm B}$ is constrained by the requirement that the mean magnetic-field energy density remain finite in the infrared ($k \to 0$), which requires $n_{\rm B}>-3$, with $n_{\rm B} \to -3$ corresponding to a scale-invariant spectrum on large scales.

The magnetic field power spectrum drops at small scales owing to dissipation in the pre-recombination era. The cut-off scale $k_{\rm max}$ is determined by the comoving Alfv\'en wave damping scale $k_{\rm max}^{-1} \sim v_A L_S$, where $v_A$ is the Alfv\'en velocity and $L_S$ the Silk damping scale. The Alfv\'en damping scale is approximately given by, $k_{\rm max} \simeq 200 \, (1 \, {\rm nG}/B_0) \, \rm Mpc^{-1}$ \cite{1998PhRvD..57.3264J}.

The principal PMF effect that we consider in this work is the generation of additional small-scale matter perturbations. Spatial fluctuations in the magnetic energy density and magnetic pressure generate Lorentz forces that act on the primordial baryonic plasma. After photon decoupling, when radiation forces become subdominant, these magnetically induced baryonic perturbations can grow gravitationally. The resulting gravitational potentials subsequently source perturbations in the dark-matter distribution, thereby contributing to the total matter perturbations \cite{Cruz2024, Sethi2005}. Assuming that the adiabatic and magnetically sourced perturbations are statistically uncorrelated, the total linear matter power spectrum can be written as,
\begin{equation}
\label{eq:total-pmf-matter-power}
P_{\mathrm m}(k,z) = D_{+}^{\,2}(z)P_{\mathrm{ad}}(k) + M^{2}(z)P_{\mathrm{pmf}}(k),
\end{equation}
where $P_{\mathrm{ad}}(k)$ is the present-day standard $\Lambda$CDM linear matter power spectrum, $D_{+}(z)$ is the conventional cosmological growth factor, $P_{\mathrm{pmf}}(k)$ is the magnetically sourced matter-power contribution and $M(z)$ describes its redshift evolution \cite{2024ApJ...972..117Z, Adi2023PMFDensity, Gopal_2003, Kim_etal_1996}. Following a refined formalism developed by \cite{2024ApJ...972..117Z}, the PMF-induced matter power spectrum $P_{\rm pmf}(k)$ can be written as,
\begin{align}
\label{eq:pmf-matter-power}
P_{\rm pmf}(k) &= \left( \frac{\alpha k}{4\pi}\right)^2 \int_{k_{\rm min}}^{k_{\rm max}} k_1^2 \ dk_1 \int_{-1}^{1} d\mu \ P_B(k_1) P_B(k_2) \ [k^2 + (k^2-2kk_1\mu)\mu^2]\,,
\end{align}
where $\mu = \hat{\mathbf k} \cdot \hat{\mathbf k}_1$ is essentially the cosine of the angle between ${\mathbf k}$ and ${\mathbf k}_1$, $k_2 = |\mathbf k-\mathbf k_1| = \sqrt{k^2+k_1^2-2kk_1\mu}$, $\alpha=f_b/(\mu_0\bar{\rho}_{\rm b,0})$, $f_b= \Omega_{\rm b}/{\Omega}_{\rm m}$, $\bar{\rho}_{\rm b,0}$ is the present day baryon density, and $\mu_0$ is the magnetic permeability of free space. For PMFs generated during the early radiation-dominated era, the infrared cutoff, $k_{\min}$, can generally be taken sufficiently small, ie., $k_{\min}\simeq0$. The scale $k_{\max}$ denotes the ultraviolet cutoff associated with the damping of the PMF through Alfv\'en-wave dissipation prior to recombination \cite{2024ApJ...972..117Z, Adi2023PMFDensity, Gopal_2003, Kim_etal_1996}. The redshift evolution, $M(z)$ is given as,
\begin{equation}
\label{eq:pmf-matter-power-evolution}
M(z) = t_0^2 \left[ \frac{9}{10} \frac{1+z_{\rm rec}}{1+z} + \frac{3}{5} \left(\frac{1+z}{1+z_{\rm rec}}\right)^{3/2} - \frac{3}{2} \right] \,,
\end{equation}
where $t_0=(6\pi G \bar{\rho}_{\rm m,0})^{-1/2}$ while, $\bar{\rho}_{\rm m,0}$ is the present day total matter density and $G$ is gravitational constant \cite{2024ApJ...972..117Z, Adi2023PMFDensity, Gopal_2003, Kim_etal_1996}. The amplitude and scale dependence of $P_{\mathrm{pmf}}(k)$ are primarily controlled by $B_0$ and $n_{\rm B}$. It can be shown that the matter power spectrum induced by magnetic fields has scale dependence, $P_{\mathrm{pmf}}(k) \propto k^{2n_{\rm B} +7}$ for $n_{\rm B} \le -1.5$. It can also be easily seen that the growth factors $D_+(z)$ and $M(z)$ normalized to the present epoch are nearly identical at low redshifts \cite{Sethi2005, Adi2023PMFDensity}.

\begin{figure}
\centering
\includegraphics[width=\textwidth]{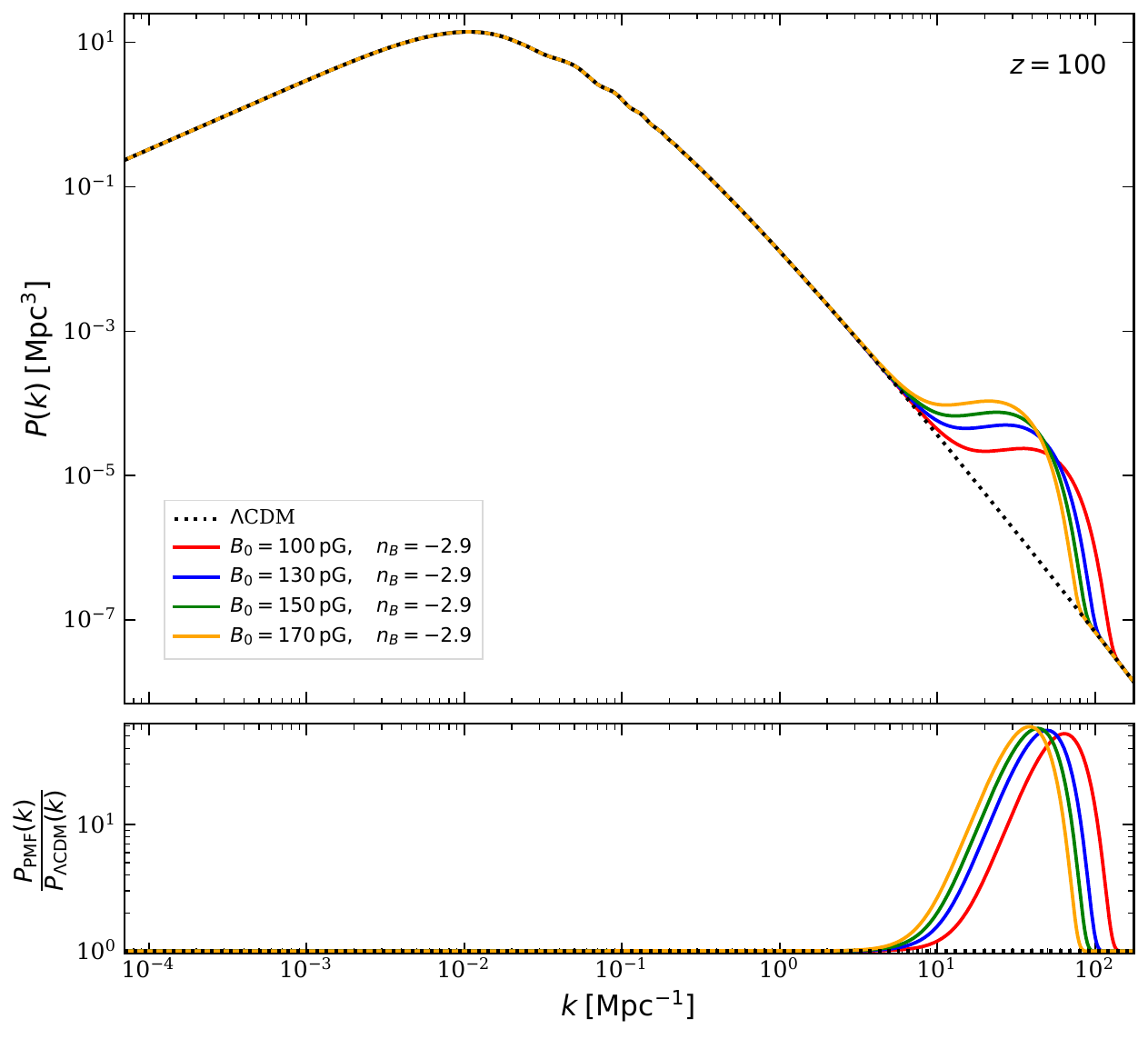}
\caption{Linear matter power spectra at $z=100$ used to initialize the simulations. The upper panel shows the matter power spectrum $P(k)$ for the $\Lambda$CDM reference (black dotted curve) and four primordial magnetic field (PMF) models with comoving magnetic field amplitudes $B_0=100$, $130$, $150$, and $170\,\mathrm{pG}$, all having the spectral index $n_B=-2.9$. The spectra closely agree on large scales ($k\lesssim 5\,\mathrm{Mpc}^{-1}$), while the PMF models exhibit enhanced power at smaller scales. The enhancement of magnitude and scale dependence vary with $B_0$, with each model showing a turnover followed by a steep decline at high wavenumbers which associates with magnetic damping. The lower panels are showing the ratio $P_{\mathrm{PMF}}(k)/P_{\Lambda\mathrm{CDM}}(k)$, where $P_m^{\mathrm{PMF}}(k)$ denotes the total matter power spectrum in the PMF cosmology, given by $P_m^{\mathrm{PMF}}(k)=P_m^{\Lambda\mathrm{CDM}}(k)+P_m^{\mathrm{mag}}(k)$, with $P_m^{\mathrm{mag}}(k)$ representing the magnetically induced contribution. This highlights the scale-dependent enhancement relative to the standard cosmological model. The ratio approaches unity on large scales, increases above unity where the magnetic contribution becomes significant, and subsequently decreases towards unity at sufficiently high wavenumbers.}
\label{fig:input-power-spectra}
\end{figure}

After recombination, magnetic-pressure gradients suppress the gravitational collapse of baryonic perturbations on scales smaller than the magnetic Jeans length, and hence the PMF-induced matter power spectrum is suppressed for scales corresponding to $k \gg k_J$. The magnetic Jeans scale can be approximately estimated using $k_J \simeq 15( 1\, {\rm nG}/B_{0}) \, \rm Mpc^{-1}$ \cite{Kim_etal_1996, Gopal_2003}. The magnetic Jeans scale does depend on $n_{\rm B}$, but the dependence is extremely weak for near scale-free models where $n_{\rm B} \simeq -3$. It is important to emphasize here that the magnetic Jeans scale $k_J$ is a time-independent quantity, given that the magnetic field redshifts as $\propto 1/a^2$, without significant distortion. It can be easily shown that the scales correspond to $k\ll k_J$. We suppress the PMF-induced matter power spectrum due to magnetic pressure by introducing a smooth exponential cutoff at the magnetic Jeans scale. The modified matter power spectrum is shown in Fig.~\ref{fig:input-power-spectra} for different values of $B_0$ and $n_{\rm B}$, along with the power spectrum for the usual $\Lambda$CDM case. It is important to note that sub-nG primordial magnetic fields have a negligible impact on the normalization of $\sigma_8$, as the PMF-induced contribution to the matter power spectrum is insignificant over the range $k\simeq0.01-0.5\,{\rm Mpc}^{-1}$ that is relevant for $\sigma_8$. In this work, we restrict our analysis to models with $n_{\rm B}$ close to $-3$, as nearly scale-invariant spectra are well motivated by several inflationary magnetogenesis scenarios and yield a relatively conservative contribution to the PMF-induced matter power spectrum.

\section{Impact of PMFs on Structure Formation and Reionization}
\label{sec:pmf_reion}
The PMF contribution in our models is only confined to small scales (see Fig.~\ref{fig:input-power-spectra}). These scales lie far beyond those accessible to the CMB. These scales are already non-linear under gravitational collapse at the redshifts of interest~\cite{Ralegankar2024ConsistentIC, Ralegankar2025nonlinearMPS}. Consequently, understanding the impact of PMFs on structure formation and reionization requires following the modified matter power spectrum into the non-linear regime, rather than relying on extrapolations from linear theory. Moreover, this evolution must be modeled within a sufficiently large cosmological volume to capture the characteristic spatial scales of the ionized regions probed by the 21-cm signal~\cite{Furlanetto2006, Pritchard2012, Mondal2016EoRCovariance}.

We address these requirements using a three-stage simulation. First, an $N$-body simulation follows the non-linear evolution of the matter density field. The dark matter haloes are then identified from the particle distribution using a friends-of-friends algorithm. These haloes are the hosts for ionizing sources. Finally, a semi-numerical excursion-set algorithm, \texttt{ReionYuga}, combines the evolved matter field and halo catalogues to construct the ionization and neutral hydrogen distributions, assuming that hydrogen exactly traces the matter field. A similar framework was previously used to study the impact of non-standard small-scale power on reionization~\cite{Sarkar2016, Das2018}.

Throughout this work, the PMF enters only through the input spectrum. As noted in Sec.~\ref{sec:pmf-review}, we do not dynamically evolve the magnetic field or the gas. We also omit thermal heating by ambipolar diffusion and decaying magnetohydrodynamic turbulence~\cite{Shiraishi2014PMF21cm, Cruz2024}. Thus, our results should be 
interpreted as quantifying the structure formation channel associated with PMFs rather than their full dynamical and thermal impact.


\subsection{\texorpdfstring{$N$-body}{N-body} Simulation}
\label{sec:simulation-processes}
We evolve the matter distribution using a collisionless gravity-only particle-mesh (PM) $N$-body code\footnote{\url{https://github.com/rajeshmondal18/N-body}}~\cite{Mondal2016EoRCovariance} in a periodic cubic volume. PM schemes trade small-scale force resolution for the ability required to track very large numbers of particles. This is the best compromise for our study, as we are interested mainly in the DM density field and the halo population, rather than the internal structure of individual haloes~\cite{HockneyEastwood1988}.

All runs adopt the same background cosmology, with $h=0.7$, $\Omega_{\rm m}=0.3$, $\Omega_{\Lambda}=0.7$, $\Omega_{\rm b}=0.0461$ and $n_{\rm s}=0.96$. The force mesh consists of $1536^{3}$ cells with a comoving spacing of $0.07\,\mathrm{Mpc}$ resulting in a box of side $107.52\,\mathrm{Mpc}$. The particles are located at every second meshpoint along each axis, and therefore the simulation contains $768^{3}$ equal-mass particles, each with a mass of $m_{\rm p}=1.12\times10^{8}\,M_{\odot}$.

Initial conditions are generated as random Gaussian fields with the linear power spectrum of each model at the starting redshift $z=100$ (Figure~\ref{fig:input-power-spectra}). All models are run with the same random seed. We displace the particles from their initial grid positions with the Zel'dovich approximation~\cite{Zeldovich1970}. The density field is produced by cloud-in-cell assignment. The potential is obtained in Fourier space by dividing the transformed density by the representation of the seven-point finite-difference Laplacian. We then recover the force via central differencing and interpolation back to the particles. Positions and velocities are advanced with a staggered leapfrog integration scheme with $\Delta a=0.004$. We keep ten separate images from redshifts 14 to 6. The mesh, particle load and phase are the same for all model runs so any differences in results are directly attributable to the input power spectrum.


\subsection{Identifying the Haloes}
\label{sec:halo-mass-function}
We identify haloes in each snapshot using a friends-of-friends (FoF) algorithm~\cite{Davis1985FoF}. It is a publicly available halo finder\footnote{\url{https://github.com/rajeshmondal18/FoF-Halo-finder}}~\cite{Mondal2015}. We use a dimensionless linking length parameter $b=0.2$, which is $0.2$ times the average inter-particle separation in comoving coordinates. For our simulation, the mean separation is $0.140\,\mathrm{Mpc}$, giving a comoving linking length of $0.028\,\mathrm{Mpc}$. Particles are connected directly or through a chain of such links which are assigned to the same group. The halo mass is simply the sum of the masses of its member particles. We retain these groups containing at least 10 particles, corresponding to a minimum halo mass of $M_{\min}\simeq1.12\times10^{9}\,M_{\odot}$. Halo masses and abundances near this threshold can be affected by particle sampling and force resolution limits\cite{Warren2006MassFunction, Knebe2011HaloFinder}. We therefore interpret the lowest-mass bins cautiously and further test the halo catalogue below.

Our choice of minimum halo mass is also physically well-motivated. For example, at $z \simeq 8$, a mass of $\sim 10^8 M_\odot$ corresponds to a virial temperature of $T_{\rm vir} \sim 10^4$\,K~\cite{BarkanaLoeb2001}. This is the HI-cooling threshold for primordial hydrogen. This threshold defines the minimum mass required for a halo to sustain efficient atomic cooling and subsequent star formation~\cite{SobacchiMesinger2013GasSuppression}. While the first stars form in metal-free minihaloes via molecular cooling at lower masses~\cite{BarkanaLoeb2001, Abel2002FirstStar}, UV photons from these stellar populations quickly photoevaporate gas from such shallow potential wells, leaving haloes with $T_{\rm vir} \ge 10^4$ K as the primary surviving sites for sustained star formation~\cite{SobacchiMesinger2013GasSuppression}. We, therefore, do not consider the possible impact of $H_2$-cooled minihaloes below the atomic cooling threshold.

Observations suggest that the reionization ends gradually around $z \sim 6$~\cite{Fan2006, Bosman2022ReionizationEnd}. Completing reionization by this redshift requires a sufficient number of ionizing photons from haloes of various masses. Decreasing $M_{\min}$ would produce more ionizing photons from smaller haloes, causing the ionization to end too early. In contrast, increasing $M_{\min}$ delays reionization. While early simulations suggested $M_{\min} \sim 10^6-10^7 M_\odot$ in the absence of metal-free Population III stars~\cite{Choudhury2008MinimumMass}, more recent results derived from observed high-redshift UV luminosity functions point toward threshold masses closer to $M_{\min} \sim 10^9 M_\odot$~\cite{MunozLoeb2011MinimumMass}. Thus our value of $M_{\min}$ as a threshold is well justified~\cite{Mondal2017EoRCovariance}.

The Press--Schechter (PS) prescription is a first-order spherical-collapse model. The Sheth--Tormen (ST) function is an ellipsoidal-collapse model calibrated against numerical simulations~\cite{ShethMoTormen2001}. An understanding of both prescriptions will shed light on the dependence of the analytical predictions on the underlying collapse model and its calibration, especially for rare, high-mass haloes. Both analytical multiplicity functions share the following general form:
\begin{equation}
\label{eq:sheth-tormen}
\frac{\mathrm{d}n}{\mathrm{d}\ln M} = \frac{\overline{\rho}_{\rm m}}{M}\, \mathcal{F}(\nu) \left| \frac{\mathrm{d}\ln\nu}{\mathrm{d}\ln M} \right|\,,
\end{equation}
where $\nu=\frac{\delta_{\rm c}}{\sigma(M,z)}$. Here $\mathrm{d}n/\mathrm{d}\ln M$ is the comoving halo number density per logarithmic mass interval, $\overline{\rho}_{\rm m}=\Omega_{\rm m}\rho_{{\rm c},0}$ is the mean comoving matter density, and $\sigma(M,z)$ is the root-mean-square variance of the linear matter density smoothed with a spherical top-hat window enclosing mass $M$. The quantities $\Omega_{\rm m}$ and $\rho_{{\rm c},0}$ denote the current matter density parameter and critical density, respectively. We adopt a constant critical collapse overdensity of $\delta_{\rm c}=1.68647$, corresponding to the Einstein--de Sitter limit, which approximates well at these high redshifts of interest. At fixed redshift, this convention gives $\mathrm{d}\ln\nu/\mathrm{d}\ln M =-\mathrm{d}\ln\sigma/\mathrm{d}\ln M$. The two multiplicity functions are given by
\begin{equation}
\label{eq:ps-multiplicity}
\mathcal{F}_{\rm PS}(\nu)=
\sqrt{\frac{2}{\pi}}\,\nu \exp\!\left(-\frac{\nu^2}{2}\right),
\end{equation}
and
\begin{equation}
\label{eq:st-multiplicity}
\mathcal{F}_{\rm ST}(\nu) = A\sqrt{\frac{2a}{\pi}}\,\nu \left[1+(a\nu^2)^{-p}\right] \exp\!\left(-\frac{a\nu^2}{2}\right),
\end{equation}
where $a=0.707$, $p=0.3$, and $A\simeq0.3222$. The parameter $A$ is fixed by $\int_0^\infty\mathcal{F}_{\rm ST}(\nu)\, \mathrm{d}\ln\nu=1$; the multiplicity function of PS satisfies the same normalization. These are the standard PS and ST forms expressed in the stated peak-height convention \cite{PressSchechter1974, ShethTormen1999, ShethMoTormen2001}.


\subsection{Generating the 21-cm Brightness Temperature Field}
\label{sec:21cm}
We construct the 3D ionization field on a $192^3$ grid by coarsening the high-resolution $1536^3$ $N$-body mesh by a factor of 8 along each axis. The resulting comoving cell size is $0.56\,\mathrm{Mpc}$. We generate ionization fields using \texttt{ReionYuga}\footnote{\url{https://github.com/rajeshmondal18/ReionYuga}}, following the framework described in \cite{Mondal2017EoRCovariance}. Simulated particles and halo distributions are interpolated onto the \texttt{ReionYuga} grid using the cloud-in-cell method. We make an assumption that hydrogen traces the dark matter distribution on the scales which are represented by the grid. Our ionization calculation only follows hydrogen but for our optical-depth calculation we treat helium separately.

Every halo hosts an ionizing source whose ionizing photon emission rate is assumed to be proportional to its mass \cite{Majumdar2014Seminumerical, Sarkar2016, Das2018}. A single dimensionless parameter, $N_{\rm ion}$, describes the effective ionizing source efficiency by combining the fraction of halo baryons converted into stars, the stellar ionizing-photon yield and the fraction of these photons that escape into the IGM. We fixed the $N_{\rm ion}=32.51$ for all our simulations. One of the reasons to do this is that the Thomson scattering optical depth value for the reference $\Lambda$CDM model is in the range of the Planck 2018 measurement $\tau_{\rm e}=0.0544\pm0.0073$~\cite{Planck2018}. The source prescription is kept constant, while the matter and halo distributions can vary, allowing us to examine the effect of PMF-modified structure formation on the timing and morphology of reionization.

We compute the ionization field using an excursion-set formalism~\cite{Bond1991ExcursionSet, Furlanetto2004HII} in the homogeneous recombination scheme of~\cite{Choudhury2009Topology, Majumdar2014Seminumerical}. The gridded matter and collapsed-halo mass fields are smoothed with spherical top-hat filters of different radius, ranging from the grid cell scale up to $R_{\max}=20\,\mathrm{Mpc}$. A grid point is flagged as fully ionized if the integrated photon budget exceeds or is equal to the hydrogen budget on any filtering scale. This condition is $N_{\rm ion}f_{\rm coll}(\boldsymbol{x},R)\ge 1$, where $f_{\rm coll}$ is the collapsed mass fraction.

To incorporate the effect of line of sight distortions, we map the neutral-hydrogen density fields on to the redshift space using the peculiar velocities of the simulation HI particles~\cite{Majumdar2013PeculiarVelocities, Majumdar2014Seminumerical, Mondal2018}. The normalized neutral-hydrogen density field is calculated as, $\eta_{\mathrm{HI}}(\boldsymbol{x},z) = \frac{\rho_{\mathrm{HI}}(\boldsymbol{x},z)} {\overline{\rho}_{\mathrm H}(z)} = x_{\mathrm{HI}}(\boldsymbol{x},z) [1+\delta_{\mathrm m}(\boldsymbol{x},z)]$, where $\rho_{\mathrm{HI}}$ is the neutral-hydrogen mass density, $\overline{\rho}_{\mathrm H}$ is the mean total hydrogen density, $x_{\mathrm{HI}}$ is the local neutral hydrogen fraction, and $\delta_{\mathrm m}$ is the matter density contrast. We calculate both the volume-averaged neutral fraction, $\overline{x}_{\mathrm{HI}}^{\,V}=\langle x_{\mathrm{HI}}\rangle_V$, and the mass-weighted neutral fraction, $\overline{x}_{\mathrm{HI}}^{\,M}=\overline{\rho}_{\mathrm{HI}}/\overline{\rho}_{\mathrm H}=\langle\eta_{\mathrm{HI}}\rangle_V$. These two measures generally differ because neutral gas is not distributed uniformly relative to the underlying matter distribution.

Finally, we construct the differential 21-cm brightness temperature field under the saturated-spin-temperature limit~\cite{Madau1997_21cm, BharadwajAli2004, Furlanetto2006, Pritchard2012}:
\begin{equation}
\label{eq:brightness-temperature}
\delta T_{\mathrm b}(\boldsymbol{x},z) = T_0(z)\,\eta_{\mathrm{HI}}(\boldsymbol{x},z),
\end{equation}
where the temperature prefactor is given by 
\begin{equation}
\label{eq:brightness-prefactor}
T_0(z) = 27\,\mathrm{mK} \left(\frac{\Omega_{\mathrm b}h^2}{0.023}\right) \left[ \frac{0.15}{\Omega_{\mathrm m}h^2} \left(\frac{1+z}{10}\right) \right]^{1/2}\,,
\end{equation}
Where $\Omega_{\mathrm b}$ and $\Omega_{\mathrm m}$ denote the present-day baryon and total-matter density parameters, and $h=H_0/(100\,\mathrm{km\,s^{-1}\,Mpc^{-1}})$.


\section{Results}
\label{sec:results}
We compare the reference $\Lambda$CDM model against four PMF models, with $B_0=100$, $130$, $150$, and $170\,\mathrm{pG}$, and a common magnetic spectral index $n_B=-2.9$. We hold the background cosmology and source prescription fixed, so this comparison isolates the effects of PMF-modified power spectra on non-linear structure formation and reionization. As noted above, our study does not include the calculation of magnetic heating or magnetohydrodynamic backreaction. From our ten stored outputs, we show six representative snapshots at $z=10.555$, $9.607$, $8.744$, $7.958$, $7.243$, and $6.592$. However, the global reionization histories and the calculation of $\tau_e$ utilize all ten snapshots.


\subsection{The Evolved Matter Power Spectrum}
\label{sec:results-matter-power}
\begin{figure}
\centering
\includegraphics[width=\textwidth]
{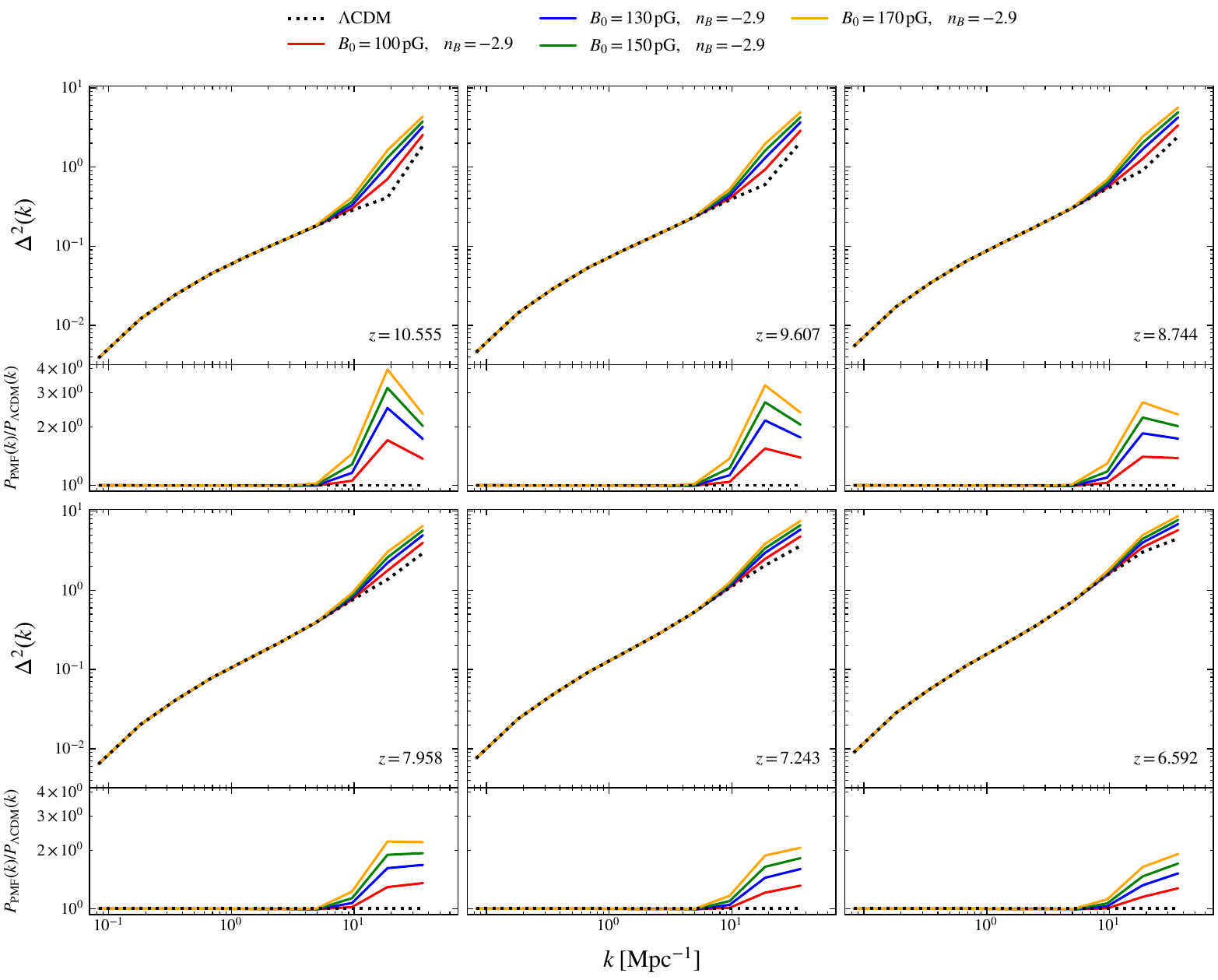}
\caption{Evolved dimensionless matter power spectra for $\Lambda$CDM and the $100$, $130$, $150$, and $170\,\mathrm{pG}$ PMF models at the six indicated redshifts. Lower panels show the ratio $P_{\rm m}^{\rm PMF}/P_{\rm m}^{\Lambda\mathrm{CDM}}$, with the dotted horizontal reference line marking unity.} 
\label{fig:output-matter-power}
\end{figure}

Figure~\ref{fig:output-matter-power} shows the dimensionless power spectrum, $\Delta_{\rm m}^{2}(k,z) = \frac{k^3P_{\rm m}(k,z)}{2\pi^2}$, of the simulated dark matter density contrast as a function of the comoving wave number $k$, alongside the ratio $P_{\rm m}^{\rm PMF}/P_{\rm m}^{\Lambda\mathrm{CDM}}$ in the lower panels. On large scales, all curves nearly coincide, whereas they separate strongly on smaller scales, consistent with the small-scale character of the PMF-induced initial power. For all the redshifts, departures from the reference model become visible at $k \sim 5 \mathrm{Mpc}^{-1}$ and grow more pronounced towards $k\sim 20\,\mathrm{Mpc}^{-1}$. The largest ratio of PMF to $\Lambda$CDM among the plotted wavenumber bins is found near $k\simeq20\,\mathrm{Mpc}^{-1}$. At this scale, the $170\,\mathrm{pG}$ model has about four times the reference power, while the $100\,\mathrm{pG}$ model has less than twice the reference power. While the overall amplitudes of all spectra increase with decreasing redshift, the reference spectrum rises more steeply with wavenumber than the PMF spectra, causing the ratios to decline with decreasing redshift.

Under a purely scale-independent linear growth factor, all spectra would be scaled by the same multiplicative factor, leaving their ratio unchanged. The measured evolution of the ratio with $z$ is therefore inconsistent with that simple linear scaling and remains qualitatively consistent with non-linear gravitational evolution that modifies the spectral shapes. However, the figure alone cannot completely distinguish this physical non-linear evolution from numerical effects at the smallest scales. In particular, force resolution, particle discreteness, density assignment, and Fourier-grid sampling can affect the highest-wavenumber bins.


\subsection{The Halo Mass Function}
\label{sec:results-halo-abundance}
We compare the simulated halo mass functions against the ST analytical prescription~\cite{ShethTormen1999, ShethMoTormen2001} in Figure~\ref{fig:hmf_fof_st_results}. The corresponding PS comparison~\cite{PressSchechter1974} is shown in Appendix~\ref{app:press-schechter} (Figure~\ref{fig:hmf_fof_ps_results}). Figure~\ref{fig:hmf_fof_st_results} show the mass function for our six selected redshifts ($z=10.555$, $9.607$, $8.744$, $7.958$, $7.243$, and $6.592$) for reference $\Lambda$CDM model and the four PMF models ($B_0=100$, $130$, $150$, and $170\,\mathrm{pG}$). Numerically calibrated mass functions are not exactly universal \cite{Reed2007MassFunction}. Their accuracy depends on the definition of the halo, the redshift, and the range of masses and cosmologies tested~\cite{Jenkins2001MassFunction, Tinker2008MassFunction, Watson2013MassFunction}. Standard calibrations do not guarantee precision for the localized enhancement and damping cutoff in our PMF spectra. We therefore use PS and ST as analytical benchmarks for the measured mass and redshift dependence, rather than as validated predictions for these PMF models. The ionizing source population used in Sec.~\ref{sec:21cm} are drawn directly from the simulated FoF catalogues, independent of these analytical curves.

The analytical curves are calculated from the input linear power spectrum for each model following the procedure described in section~\ref{sec:halo-mass-function}, without fitting their amplitudes to the FoF measurements. The lower panels show the percentage difference between the measured FoF abundance and the corresponding ST prediction averaged over each mass bin. Positive values indicate an excess of FoF haloes over ST, while negative values indicate a deficit. Note that these residuals compare each model against its corresponding ST prediction, rather than showing the difference between a PMF model and $\Lambda$CDM.

For example, at $z=10.555$, the PMF catalogues contain more low-mass haloes than $\Lambda$CDM, with abundances scaling monotonically for larger magnetic field amplitudes. This separation decreases with increasing mass, consistent with additional small-scale power. Because large-scale smoothing suppresses the high-wavenumber contribution, the high-mass analytical curves remain nearly coincident. At lower redshifts, overall abundances increase, and bins meeting our count threshold extend to larger masses. While the relative low-mass separation diminishes as $z$ decreases, the PMF excess persists. At the high-mass end, however, smaller model differences and sparse counts increase the measurement uncertainty. Finite-volume sample variance can contribute additional uncertainty to halo abundances beyond Poisson counting \cite{HuKravtsov2003SampleVariance}.

The ST model is rather good for the mass dependence and the high mass fall-off, but differences not so apparent on logarithmic axes appear in the residuals. At $z=10.555$ and $9.607$, many low and intermediate mass bins are lower than the ST prediction by tens of percent. Some intermediate mass bins agree better at later epochs but low mass deficits persist. Thus the agreement is not uniform across the mass scales. High-mass bins are sparse and have larger fractional counting uncertainties and occasional positive excursions. For example, at $z=7.958$ some high-mass PMF residuals are $\sim 100\%$, indicating that ST predicts about twice the bin-average abundance. The physical significance of these isolated high-mass points is low due to the small counts, omitted sample variance and uncertainty in the analytical prediction.

These low-mass deficits are likely due to particle and force resolution effects, halo finding biases or limitations of the ST formulation. This comparison alone does not separate out these effects, nor does it establish numerical convergence. Empirical calibrations of the mass function are sensitive to the halo definition and the particular simulations used to obtain them~\cite{Jenkins2001MassFunction}. This dependence on the analytical prescription is further emphasized by the PS comparison in Appendix~\ref{app:press-schechter}. Both prescriptions give qualitative benchmarks. Our simulated catalogs provide the populations of ionizing sources for the reionization simulation.

\begin{figure}
\centering
\includegraphics[width=\textwidth]
{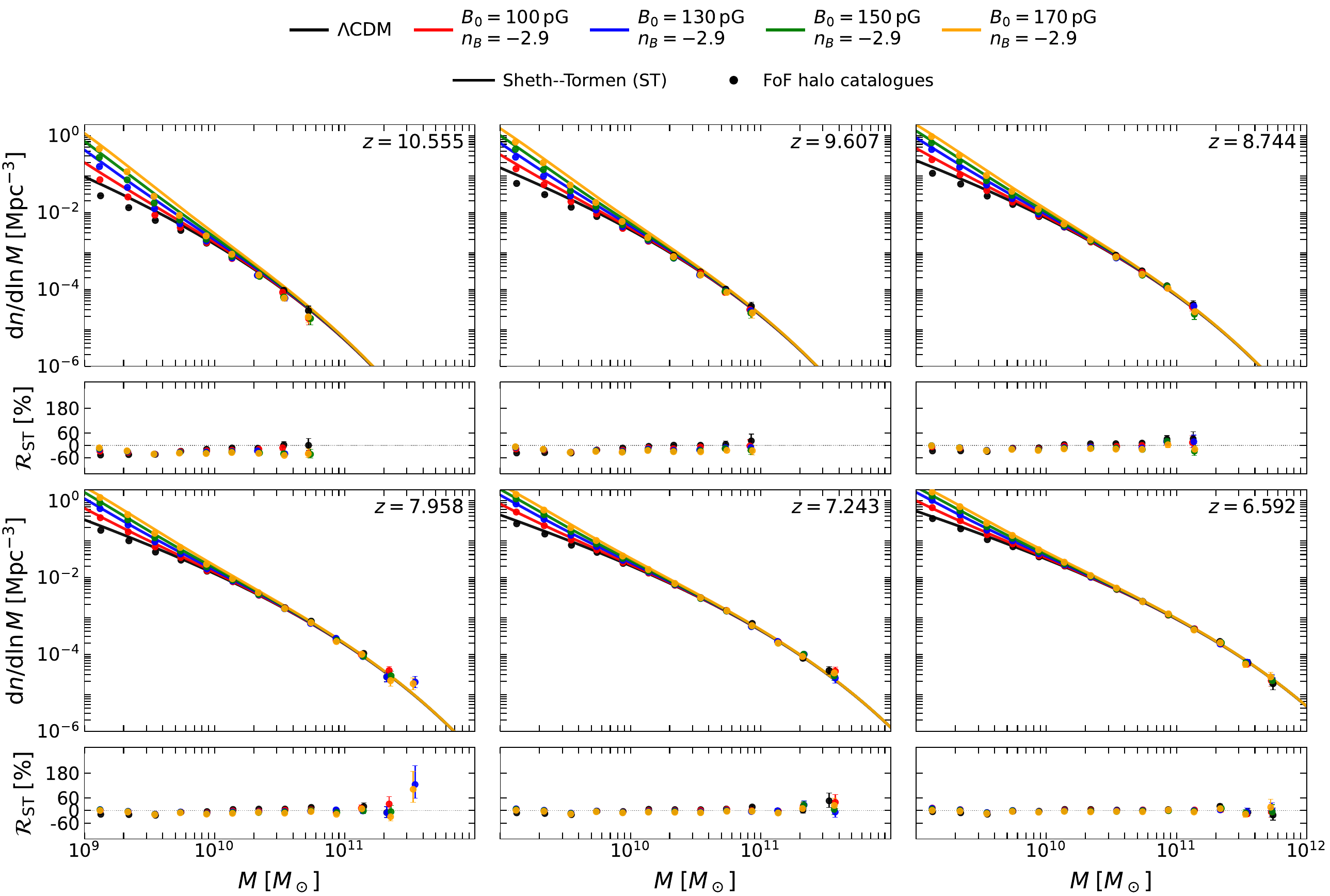}
\caption{FoF halo mass functions (points) and ST predictions (solid curves) for $\Lambda$CDM and the $100$, $130$, $150$, and $170\,\mathrm{pG}$ models, with $n_B=-2.9$ for the PMF runs, at the six indicated redshifts. Masses are in $M_\odot$, and abundances are in comoving $\mathrm{Mpc}^{-3}$ per unit $\ln M$.}
\label{fig:hmf_fof_st_results}
\end{figure}
%


\subsection{HI 21-cm Brightness Temperature Maps}
\label{sec:results-brightness-maps}
\begin{figure}[p]
\centering
\includegraphics[
width=\textwidth,
height=0.78\textheight,
keepaspectratio]{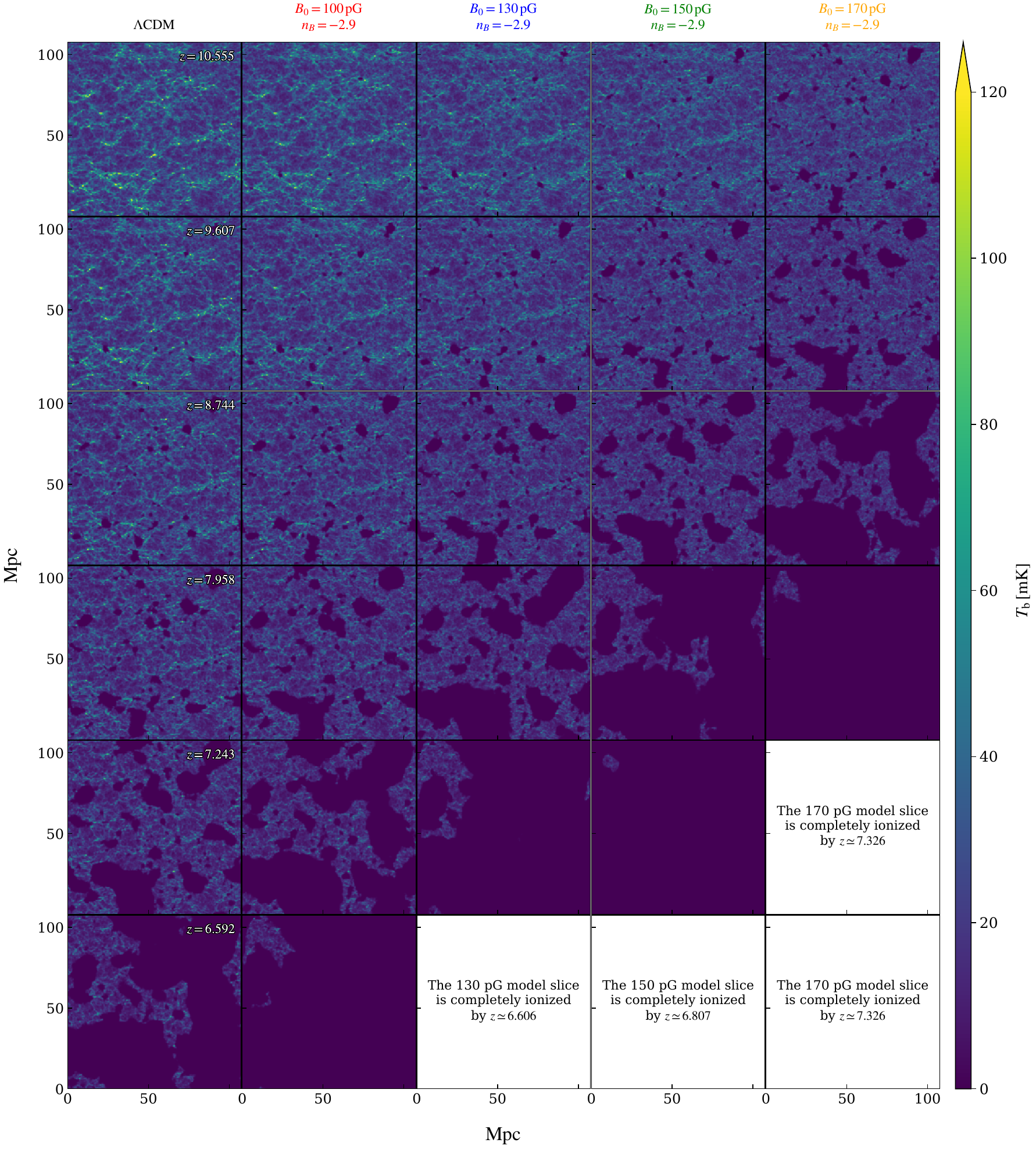} \caption{Redshift-space 21-cm brightness-temperature slices for $\Lambda$CDM and the $100$, $130$, $150$, and $170\,\mathrm{pG}$ models (columns). Rows show six redshifts $z=10.555$, $9.607$, $8.744$, $7.958$, $7.243$, and $6.592$ from top to bottom. Each slice spans $107.52\,\mathrm{Mpc}$ in comoving length and shares the same colour scale in $\mathrm{mK}$. White panels indicate zero signal across the entire slice.}
\label{fig:brightness-maps}
\end{figure}

Figure~\ref{fig:brightness-maps} shows central two-dimensional slices through the redshift-space 21-cm brightness-temperature fields constructed using equations~\eqref{eq:brightness-temperature} and~\eqref{eq:brightness-prefactor}. The maps incorporate line-of-sight peculiar velocity effects through the remapping of the neutral hydrogen field. On the common colour scale, blue--green and yellow indicate progressively stronger 21-cm emission, while dark purple includes regions of zero signal.

At $z=10.555$, all models exhibit a qualitatively similar large-scale pattern, but higher-amplitude PMF models show more conspicuous dark ionized regions. These small ionized regions expand and coalesce more rapidly with increasing magnetic field amplitude in PMF models, as seen clearly at $z=9.607$ for $170\,\mathrm{pG}$. This progression is consistent with ionized regions growing around clustered sources and merging as their combined photon budget increases~\cite{Furlanetto2004HII}.For a fixed source prescription, the larger early halo population in the PMF models speeds up reionization with increasing magnetic field amplitude. By $z=7.958$, the remaining neutral regions in the $150$ and $170\,\mathrm{pG}$ slices are small and fragmented, while $\Lambda$CDM and $100\,\mathrm{pG}$ retain considerably more neutral hydrogen. The fixed redshift comparison of these slices results in combined differences in the reionization stage and the distribution of ionized bubble sizes.

The white panels show that the HI in these specific slices is totally ionized. The $170\,\mathrm{pG}$ slice has zero signal at $z=7.243$, while the global volume-averaged neutral fraction is $\overline{x}_{\rm HI}^{\,V}=2.81\times10^{-5}$. Similarly, the $130$ and $150\,\mathrm{pG}$ slices have a zero signal at $z=6.592$, while their global neutral fractions are $\overline{x}_{\rm HI}^{\,V}=4.95\times10^{-3}$ and $5.09\times10^{-6}$, respectively. These non-zero averages confirm that neutral gas remains elsewhere in the volume. However, for $170\,\mathrm{pG}$ at $z=6.592$, the entire HI cube is completely ionized ($\overline{x}_{\rm HI}^{\,V}=0$) by $z\simeq7.326$. Because these comparisons are presented at fixed redshifts, differences in morphology reflect changes in density and source distributions, as well as shifts in the global ionization stage. Comparisons at a common global neutral fraction provide a complementary approach \cite{Sarkar2016, Das2018}.
%


\subsection{The EoR 21-cm Brightness Temperature Power Spectrum}
\label{sec:results-hi-power}
\begin{figure}
\centering
\includegraphics[width=\textwidth]
{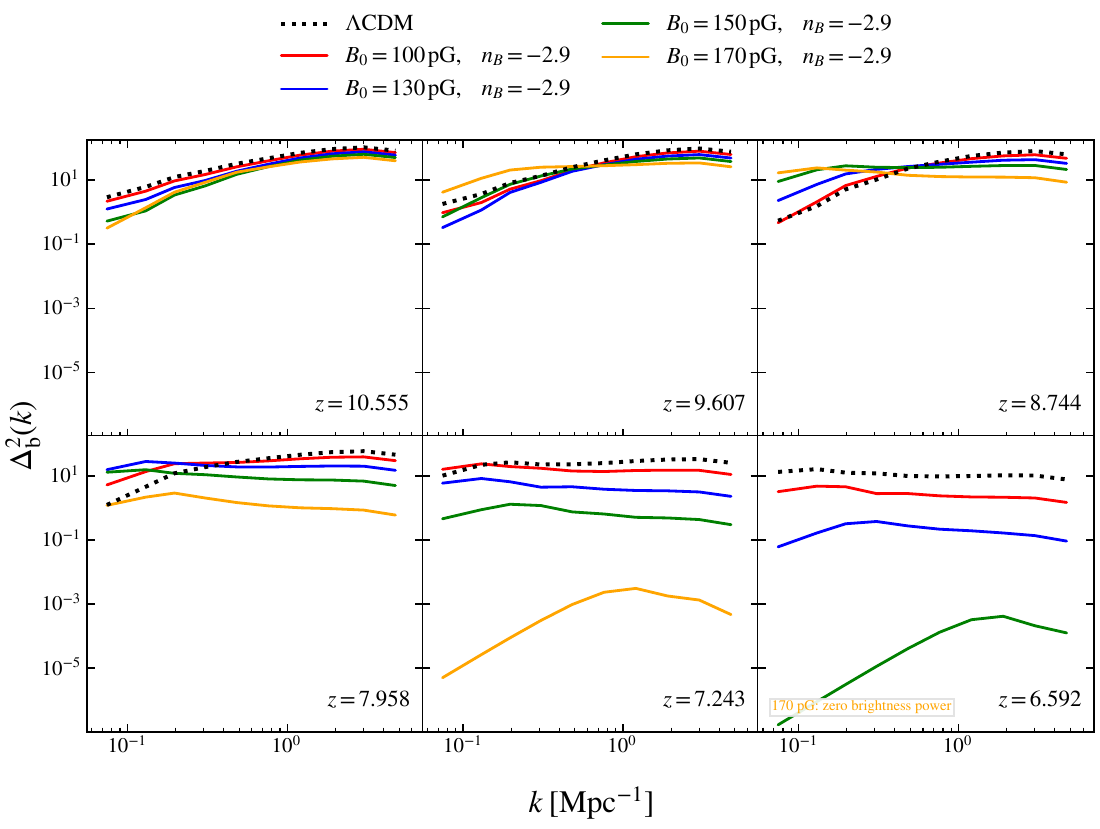}
\caption{Spherically averaged redshift-space 21-cm brightness-temperature power spectra in unit of $\mathrm{[mK]}^2$ for $\Lambda$CDM and the $100$, $130$, $150$, and $170\,\mathrm{pG}$ models at the six indicated redshifts.}
\label{fig:hi-power-spectra}
\end{figure}

Figure~\ref{fig:hi-power-spectra} shows the spherically averaged redshift-space 21-cm brightness temperature power spectrum, $\Delta_{\rm b}^{2}(k)$, for the $\Lambda$CDM reference and the four PMF models. The spectra are expressed in $\mathrm{[mK]}^{2}$ units and incorporate line-of-sight peculiar velocity distortions through the H\,I remapping described in Section~\ref{sec:21cm}. At $z=10.555$, all power spectra rise with wavenumber over most of the displayed $k$-range. The PMF curves lie below or close to the $\Lambda$CDM reference, with their separation most apparent at large scales (low $k$). By $z=9.607$, the $170\,\mathrm{pG}$ model has $\sim 133\%$ more power than the reference at $k\simeq0.075\,\mathrm{Mpc}^{-1}$, while at $k\simeq4.73\,\mathrm{Mpc}^{-1}$ its power drops $\sim 66\%$ below $\Lambda$CDM. This scale dependence is consistent with the advanced ionization state in the $170\,\mathrm{pG}$ model. In an inside-out reionization scenario, reionization around overdense sources changes the spatial correlation between density and neutral fraction \cite{Majumdar2013PeculiarVelocities}. The difference between ionized regions and the surrounding neutral gas can enhance large scale fluctuations \cite{Furlanetto2004HII}.

For $z=8.744$, the stronger magnetic field models have more large scale power than $\Lambda$CDM, with their spectra flattening and dropping below the reference at higher $k$. The models are at different stages of reionization at any given redshift (e.g. $\overline{x}_{\rm HI}^{\,V} \approx 0.900$ in the reference and $\overline{x}_{\rm HI}^{\,V} \approx 0.280$ for $170\,\mathrm{pG}$), so these power spectra reflect coupled differences in both the underlying structure and the ionization state~\cite{Sarkar2016}. 

At $z=7.958$ the $130$ and $150\, \mathrm{pG}$ models have more power than the reference on the large scales, while the $170\,\mathrm{pG}$ spectrum falls off significantly over all $k$. At this redshift, the volume-averaged neutral fractions for the $130$, $150$, and $170\,\mathrm{pG}$ models are $\overline{x}_{\rm HI}^{\,V}\simeq0.428$, $0.177$, and $0.0265$, respectively, as compared to $\simeq0.786$ in the reference. By $z=7.243$, power in the $170\,\mathrm{pG}$ model has decreased sharply across the entire $k$-range. This weak signal is consistent with the very small neutral fraction remaining in the volume ($\overline{x}_{\rm HI}^{\,V}\simeq2.81\times10^{-5}$). By $z=6.592$, the $130$ and $150\,\mathrm{pG}$ models also have substantially less power than the reference, while $\overline{x}_{\rm HI}^{\,V} = 0$ for $170\, \mathrm{pG}$ model. Consequently, the brightness-temperature field and its power vanish for $170\,\mathrm{pG}$ model at this final snapshot, as indicated by the annotation in the figure.

The evolving power spectrum reflects matter clustering, fluctuations in the neutral fraction and their spatial correlation. Line-of-sight peculiar velocities redistribute the signal in redshift space and further modify its power \cite{Majumdar2013PeculiarVelocities, Majumdar2014Seminumerical}. During the later stages of reioniztion, models with higher $B_0$ can have weak absolute brightness-temperature power despite large fractional H\,I-density fluctuations because their mean neutral hydrogen density is small. The corresponding fractional H\,I-density power spectra are shown in Appendix~\ref{app:hi-power}.
%


\subsection{Reionization History}
\label{sec:results-neutral-fraction}
\begin{figure}
\centering
\includegraphics[width=\textwidth]
{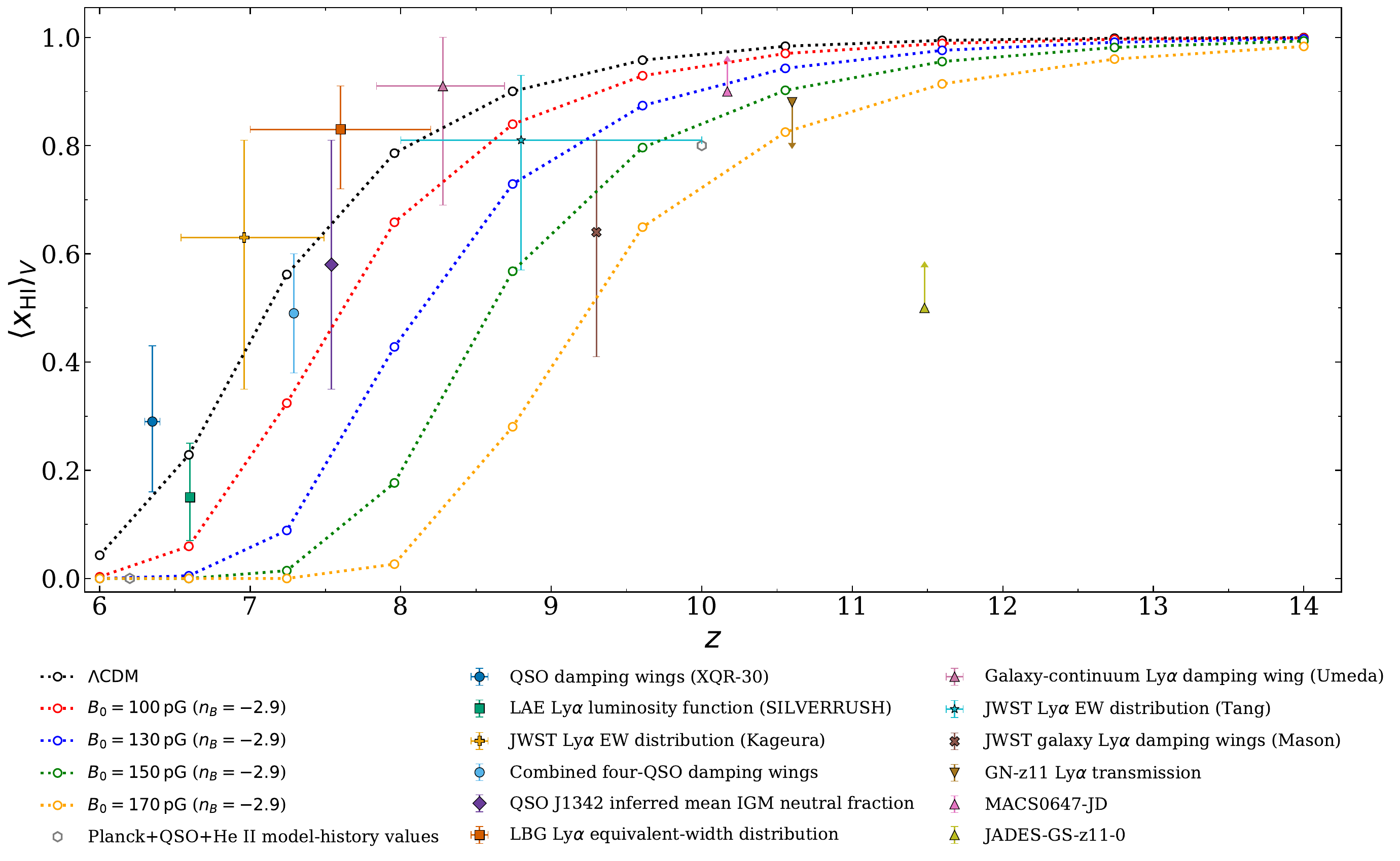}
\caption{Volume-averaged neutral fractions for the $\Lambda$CDM reference and the four PMF models. Open circles mark simulation snapshots, connected by dotted lines. Coloured symbols show observational inferences, with arrows indicating one-sided bounds and horizontal bars indicating redshift ranges. Grey open symbols represent approximate values from a published model reionization history.}
\label{fig:neutral-fraction-constraints}
\end{figure}

Figure~\ref{fig:neutral-fraction-constraints} shows the volume-averaged neutral fraction, $\overline{x}_{\rm HI}^{\,V}$, for the $\Lambda$CDM reference and the four PMF models. Increasing $B_0$ accelerates the ionization of the Universe and pushes the end to higher redshift. This is consistent with the enhanced early halo population under the fixed ionizing-source prescription.

We define the reionization midpoint $z_{50}$ by $\overline{x}_{\rm HI}^{\,V}(z_{50})=0.5$ and estimate it by linear interpolation between the two snapshots that bracket this value. The reference model gives $z_{50}\simeq7.12$, while the $100$, $130$, $150$, and $170\,\mathrm{pG}$ models give $z_{50}\simeq7.62$, $8.15$, $8.61$, and $9.26$, respectively. Defining the shift relative to the reference as $\Delta z_{50}=z_{50}^{\rm PMF}-z_{50}^{\Lambda\mathrm{CDM}}$, we obtain $\Delta z_{50}\simeq0.50$, $1.02$, $1.49$, and $2.14$, respectively. The offsets are calculated before rounding, and the estimates are limited by snapshot spacing.

Comparison with observations depends on the redshift, the observational probe, and the modeling assumptions used to infer the neutral fraction. At $z=6.6$, only $\Lambda$CDM lies within the SILVERRUSH Ly$\alpha$ luminosity-function constraint ($\overline{x}_{\rm HI}^{\,V}=0.15^{+0.10}_{-0.08}$)~\cite{Umeda2025SILVERRUSH}. At $z=6.35$, all five histories lie below the XQR-30 quasar damping-wing interval ($0.29^{+0.14}_{-0.13}$), though the reference is close to its lower edge~\cite{Greig2024XQR30}. At $z=10.17$, the volume-averaged neutral fractions of the $\Lambda$CDM, $100$, and $130\,\mathrm{pG}$ histories lie above the MACS0647-JD lower limit ($x_{\rm HI}^{\,V}>0.90$), which assume no damped Ly$\alpha$ absorber contribution~\cite{Hsiao2024MACS0647JD}.

The remaining quasar and galaxy measurements span various stages of reionization~\cite{Greig2022QsoDampingWings, Kageura2025LyaCensus, Bolan2022LBG, Umeda2024JWSTDampingWings, Tang2024JWSTLya, Mason2026JWSTDampingWings, Bruton2023GNz11}. Agreement varies across redshifts and probes, and inferred global fractions remain model-dependent. For example, the J1342+0928 result uses mapping local damping-wing constraints to the mean IGM neutral fraction~\cite{Kist2026LocalTopology}.

The JADES-GS-z11-0 data is retained as the original, model-dependent inference of Curtis-Lake et al.~\cite{CurtisLake2023JADES}. The plotted redshift was derived under the assumption of a fully neutral IGM. Subsequent spectroscopy favored a revised redshift and indicated potential damped Ly$\alpha$ contamination~\cite{Hainline2024JADES}. The gray open symbols are approximate values from a data-constrained model history and are not independent measurements~\cite{Mitra2018PlanckQuasars}.

No single history satisfies all constraints plotted versus redshift. However, this comparison is only descriptive and not a joint exclusion test. This is because individual inferences utilize different modeling assumptions, selection functions, and confidence conventions. Crucially, the accelerated reionization in stronger PMF models is contingent on the fixed source prescription. Recalibrating the source efficiency separately for each model could adjust these histories. Quantitative PMF constraints therefore require marginalizing over astrophysical uncertainties in addition to observational likelihoods.
%

\subsection{Thomson Optical Depth}
\label{sec:results-optical-depth}
The Thomson optical depth measures the integrated line-of-sight free-electron column density weighted by the Thomson scattering cross-section. We calculate this using the standard relation~\cite{Cruz2024}
\begin{equation}
\label{eq:thomson-optical-depth-general}
\tau_{\rm e} = c\sigma_{\rm T} \int_0^{z_{\rm start}} \frac{\overline{n}_{\rm e}(z)} {(1+z)H(z)} \,\mathrm{d}z \, ,
\end{equation}
where $\tau_{\rm e}$ is dimensionless, $c$ is the speed of light, $\sigma_{\rm T}$ is the Thomson scattering cross-section and $\overline{n}_{\rm e}(z)$ is the mean free-electron number density. To calculate $\tau_{\rm e}$ from our simulated reionization histories, we expand Equation~\ref{eq:thomson-optical-depth-general} as:
\begin{equation}
\label{eq:thomson-optical-depth}
\begin{aligned}
\tau_{\rm e} ={}& \frac{c\sigma_{\rm T}\Omega_{\rm b}\rho_{{\rm c},0}} {m_{\rm proton}H_0} \Bigg[\left(1-\frac{Y_{\rm p}}{2}\right) \int_0^{z_{\rm HeIII}} 
\frac{[1-\overline{x}_{\rm HI}^{\,M}(z)](1+z)^2}{E(z)} \,\mathrm{d}z \\
    &\quad+ \left(1-\frac{3Y_{\rm p}}{4}\right) \int_{z_{\rm HeIII}}^{z_{\rm start}} \frac{[1-\overline{x}_{\rm HI}^{\,M}(z)](1+z)^2}{E(z)}\,\mathrm{d}z \Bigg] \,,
\end{aligned}
\end{equation}
where $E(z)\equiv\frac{H(z)}{H_0} = \sqrt{\Omega_{\rm m}(1+z)^3+\Omega_\Lambda}$, and $m_{\rm proton}$ is the physical proton mass, and $Y_{\rm p}$ is the primordial helium mass fraction. The parameters $\rho_{{\rm c},0}$, $\Omega_{\rm b}$, $\Omega_{\rm m}$, and $\Omega_\Lambda$ denote the present-day critical density, baryon, total-matter, and dark energy density parameters, respectively. Radiation density is neglected. $\overline{x}_{\rm HI}^{\,M}$ represents the hydrogen-mass-weighted neutral fraction. Mass weighting is required because optical depth depends on the mean electron number density, which includes the correlation between density and ionization state, rather than on the ionized volume fraction alone.

\begin{figure}
\centering
\includegraphics[width=\textwidth]{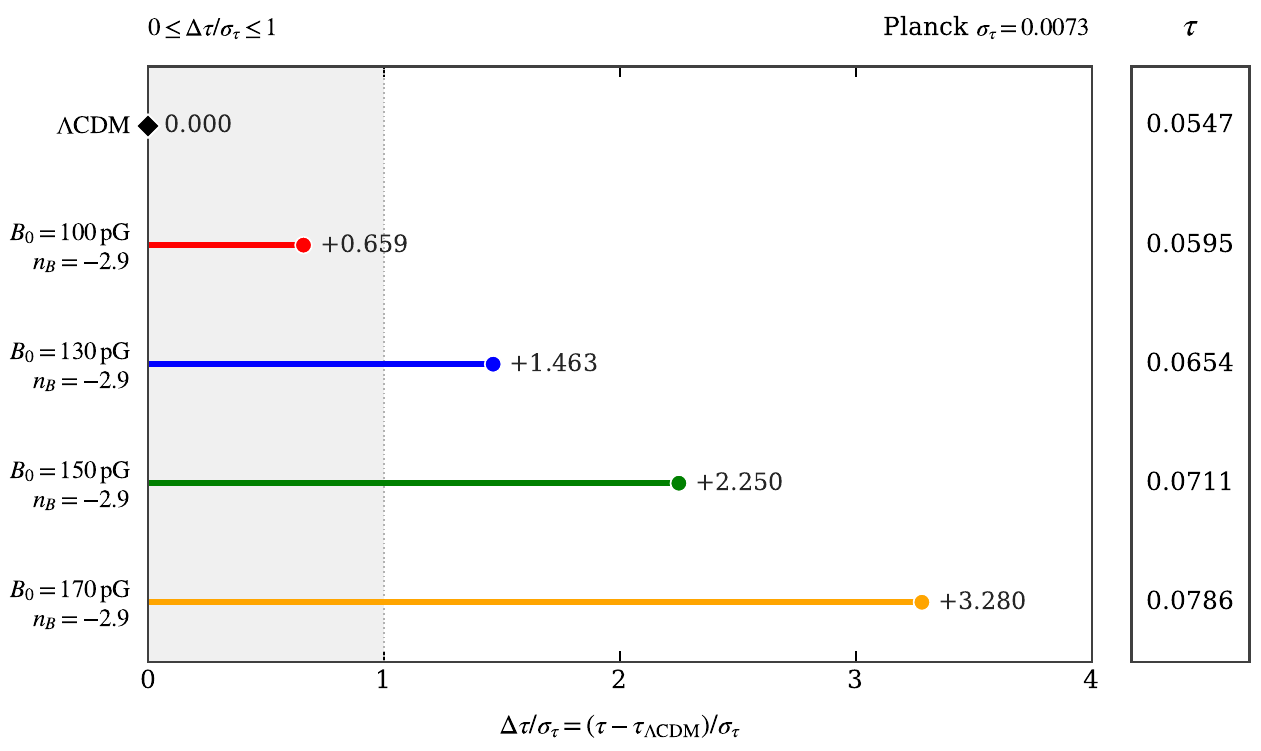}
\caption{Calculated Thomson optical depths and their offsets from the $\Lambda$CDM reference in units of $\sigma_\tau=0.0073$. Coloured segments show offsets, not uncertainty bars. The shaded interval marks an increase of up to $1\sigma_\tau$ above the reference.}
\label{fig:optical-depth}
\end{figure}

Thomson optical depth systematically increases with magnetic field for the PMF models, as shown in Figure~\ref{fig:optical-depth}. The reference $\Lambda$CDM yields $\tau_{\rm e}=0.0547$, rising monotonically to $0.0786$ for $170\,\mathrm{pG}$ model. For comparison, we take the characteristic uncertainty $\sigma_\tau = 0.0073$ from the Planck 2018 TT,TE,EE+lowE+lensing constraint, derived under a $\tanh$ reionization history parameterization~\cite{Planck2018}. Note that $\sigma_\tau$ serves purely as a reference baseline rather than an estimated error for the simulated histories. For $B_0=100$, $130$, $150$, and $170\,\mathrm{pG}$ at $n_B=-2.9$, the offsets with respect to $\Lambda$CDM baseline are $S_\tau=[\tau_{\rm e}(B_0)-\tau_{\rm e}^{\Lambda{\rm CDM}}] /\sigma_\tau = 0.659$, $1.463$, $2.250$, and $3.280$, respectively. These offsets are calculated relative to our simulated reference rather than the observational central value $0.0544$. The shaded $0$--$1$ band serves as a visual reference rather than Planck's posterior interval. These offsets are not formal exclusion significances either, but relative sensitivity measures.

The increase in $\tau_{\rm e}$ tracks the shift of $z_{50}$ toward higher redshifts, though the two statistics capture different information. While $z_{50}$ marks a discrete point in the reionization history, $\tau_{\rm e}$ integrates the mass-weighted ionized history along the line-of-sight with redshift-dependent density and path-length weighting. Enhanced early ionization increases the free-electron density along the CMB photon trajectory, increasing the scattering optical depth. Within our fixed source prescription, the PMF-enhanced small-scale clustering, increased early low-mass halo abundance, and accelerated ionization provide a consistent interpretation of the optical-depth trend. 
%

\section{Discussion and Conclusions}
\label{sec:discussion-conclusions}
We have investigated how PMFs are modifying the initial matter power spectrum which in return will influence non-linear structure formation and reionization, using $N$-body simulations coupled to a semi-numerical ionization model \texttt{ReionYuga}. We have kept the cosmology and ionizing source prescription fixed across all models, our three stage process connects the changes in the initial clustering to the evolved matter distribution, the halo population, and the resulting reionization. Our calculation deliberately isolates the structure-formation contribution, omitting subsequent magnetic-field evolution and dissipative heating.

We find that the spectrum of the evolved matter shows a scale-dependent excess over the $\Lambda$CDM reference, but the relative enhancement decreases toward lower redshift. The contrast between the models is modified by non-linear gravitational evolution without erasing the imprint of initial conditions on small scales. Thus, the higher number of early low mass source haloes results in a faster ionization given a common source efficiency prescription. This is because the physical ionization response depends on the integrated small-scale power over halo-forming scales, not on the power at a single wavenumber.

For the main sequence with $n_B=-2.9$ and $B_0=100$--$170\,\mathrm{pG}$, the reionization midpoint is earlier by $\Delta z_{50}\simeq0.50$--$2.14$ relative to $\Lambda$CDM. Similarly, the Thomson optical depth increases from $\tau_{\rm e} \simeq 0.0547$ in the reference to $0.0595$--$0.0786$ in PMF models. We adopt $N_{\rm ion}=32.51$ to bring the reference optical depth close to the Planck 2018 value~\cite{Planck2018}. The optical depth and reionization midpoint are sensitive to different things. $z_{50}$ is a discrete point in the reionization history, while $\tau_{\rm e}$ is an integral of the mean-weighted free-electron density over cosmic time. The simulation-relative offsets are in the range of $\sim 0.66$--$3.28\sigma_\tau$ with the Planck uncertainty $\sigma_\tau=0.0073$~\cite{Planck2018}. These numbers are indicators of sensitivity, not formal statistical exclusion limits. 

We have also done a supplementary exploration at a steeper spectral index, $n_B=-2.995$ (Appendix~\ref{app:pmf-large-amplitude}) which reveals a flattening of the Thomson optical depth response. This difference shows that the structure formation and reionization response is not defined by field amplitude alone. This phenomenon is linked to the amplitude-dependent magnetic Jeans length~\cite{Kim_etal_1996, Gopal_2003, Cruz2024}, which shifts the characteristic scale of the induced matter contribution.

Existing observational constraints are used to put the range of magnetic field amplitudes explored here in context. For a non-helical field, the Planck 2015 analysis obtained a $95\%$ upper limit of $B_{1\,\mathrm{Mpc}}<4400\,\mathrm{pG}$. For nearly scale-invariant fields, the corresponding limit was $2000\,\mathrm{pG}$, tightening to $900\,\mathrm{pG}$ when the magnetic contribution to the ionization history was included \cite{PlanckPMF2015}. Reionization observations can yield more stringent, but model-dependent, constraints. For example, Pandey et al.\ obtained $B_0<358\,\mathrm{pG}$ for $n_B=-2.95$ and $B_0<120\,\mathrm{pG}$ for $n_B=-2.9$, at $95\%$ confidence \cite{Pandey2015}. The latter limit is within the amplitude range explored here. However, a direct comparison requires that the field definitions, smoothing scales, and model assumptions be matched. Our higher-amplitude models therefore illustrate the response of structure formation and reionization to stronger PMFs; their inclusion does not by itself establish their observational viability.

Our results are qualitatively consistent with previous semi-analytical and excursion-set studies (e.g., Cruz et al.\ \cite{Cruz2024}). In this regard, our approach advances beyond theirs by following the non-linear gravitational evolution and assigning ionizing sources to the resolved halo distribution. Our larger simulation box captures large-scale distributions and observable statistics, while being complementary to high-resolution radiation-hydrodynamic simulations such as SPHINX-MHD~\cite{Katz2021SPHINXMHD} which model small-scale radiative and magnetic feedback in very small volumes. Our calculation, however, does not evolve the gas or magnetic field and cannot capture the dynamical back reaction of these or the associated feedback processes. Thus, the two methods give complementary information on the structure-formation channel and the coupled radiation--MHD problem. Comparisons to radiative-transfer simulations also show that agreement of approximate reionization methods depends on the spatial scale and statistic considered \cite{Zahn2011ReionizationComparison}.

The scientific value of our analysis is that we follow a proposed magnetic signature through a number of related steps, rather than judging it on a single observable. Improved initial clustering, a modified halo population and earlier ionization provide a consistent interpretation within the model adopted. At the same time, the different responses of the two spectral-index sequences suggest that an amplitude-only description might miss significant changes in the supplying scales of the sources. These links define which parts of the structure formation calculation need to be probed in interpreting reionization as a probe of PMFs. For these reasons, an observed early ionization history does not uniquely imply a PMF in the absence of ruling out astrophysical variations.

In near future work we will include magnetic-field evolution coupled to gas dynamics and the additional non-ideal and dissipative processes necessary to assess magnetic heating. Coupling these calculations with radiation and spin-temperature evolution would enable comparisons with radiation-MHD studies \cite{Katz2021SPHINXMHD}. We also plan to use the Markov chain Monte Carlo (MCMC) method to forecast constraints on PMF parameters from 21-cm observations with e.g., LOFAR, HERA and SKA-Low, while jointly varying astrophysical parameters \cite{Greig2015_21CMMC}. Combining integrated optical-depth constraints with evolved neutral-fraction histories and spatial 21-cm statistics will provide a robust framework for testing whether primordial magnetism can be reliably distinguished from variations in standard astrophysical sources.

\section{Acknowledgments}
\label{acknowledgments}
KP acknowledges financial support from Vellore Institute of Technology through its Seed Grant (No. SG20230039, 2023) and also thank the Centre for Technical Support at VIT Vellore for providing the required computational facilities. RM was supported by the NITC FRG Seed Grant (NITC/PRJ/PHY/2024-25/FRG/12).


\clearpage
\appendix

\section{Press--Schechter Mass Function}
\label{app:press-schechter}
\begin{figure}
\centering
\includegraphics[width=\textwidth]
{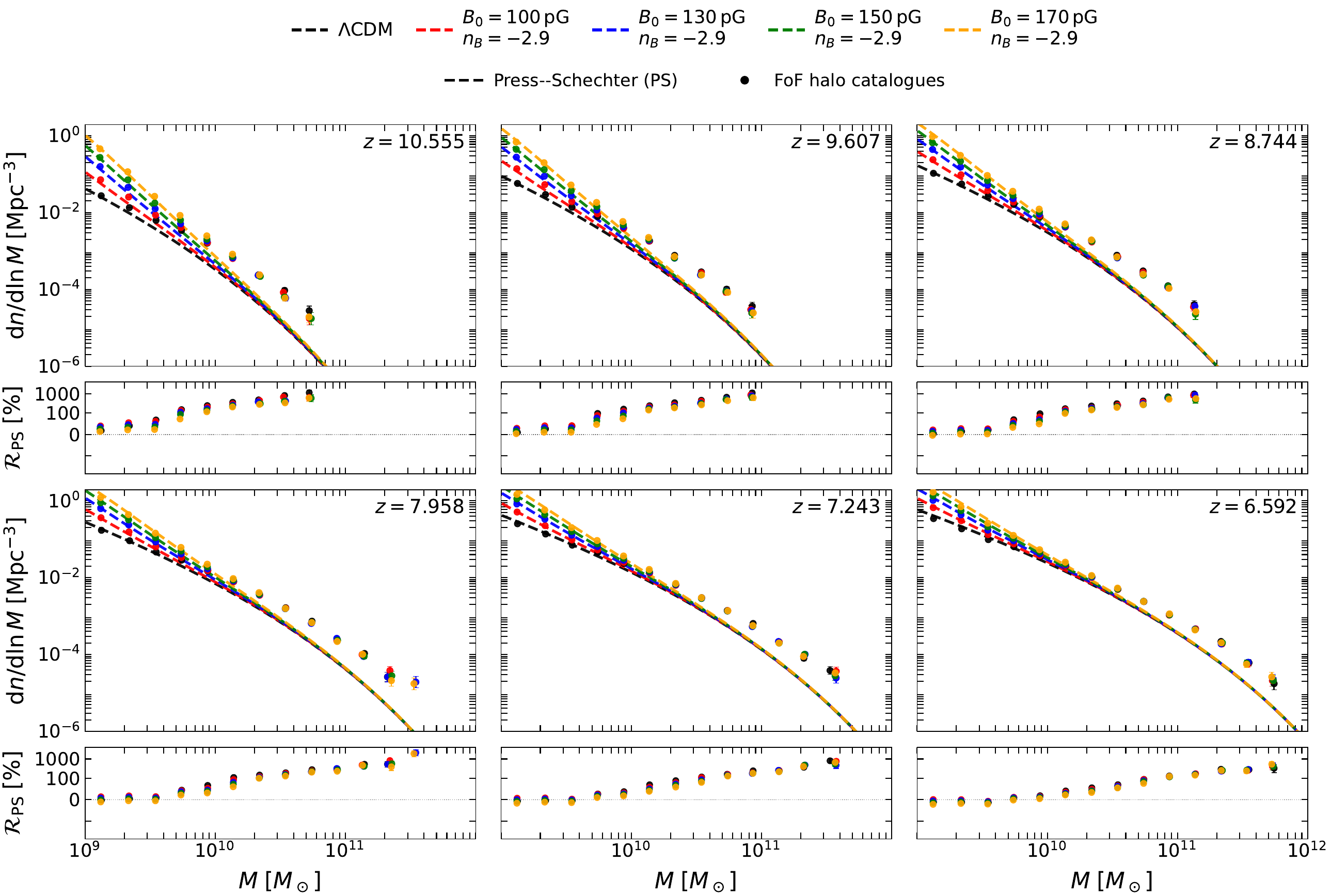}
\caption{ FoF halo mass functions compared with Press--Schechter predictions (dashed curves) for the same five models and six snapshots as in figure~\ref{fig:hmf_fof_st_results}. The residual axes use a symmetric logarithmic scale, linear within $|\mathcal R_{\rm PS}|\leq100\%$. This differs from the linear ST residual axes; the visual sizes of the residuals should therefore not be compared without accounting for the axis scales.}
\label{fig:hmf_fof_ps_results}
\end{figure}

Figure~\ref{fig:hmf_fof_ps_results} compares FoF catalogues with Press--Schechter (PS) prescription \cite{PressSchechter1974}. The models, snapshots, mass bins, halo-membership threshold, and minimum bin population are identical to those used for the Sheth--Tormen (ST) comparison in Section~\ref{sec:results-halo-abundance}. The PS predictions are calculated from the input linear spectrum using the same growth convention, without fitting or renormalizing the curves to the catalogue measurements.

PS reproduces the overall decline of abundance with increasing mass, but increasingly underpredicts the populated high-mass tail. The resulting residuals reach several hundred percent and, in some bins, values of order $10^3\%$. A residual of $900\%$, for example, means that the abundance averaged in the measured bin is ten times the prediction of PS.

The difference between PS and ST is particularly pronounced for rare haloes. For $\nu=\delta_{\rm c}/\sigma(M,z)\gg1$, the multiplicity function of PS contains the exponential factor $\exp(-\nu^2/2)$, while ST contains $\exp(-a\nu^2/2)$ with $a=0.707$~\cite{ShethTormen1999}. In ST mass function weaker exponential suppression explains its larger rare-halo prediction and helps account for its closer agreement with the measured high-mass abundances. This does not establish ST as a precision calibration for PMF-modified spectrum.


\section{ H\,I Power Spectrum}
\label{app:hi-power}
\begin{figure}
\centering
\includegraphics[width=\textwidth]
{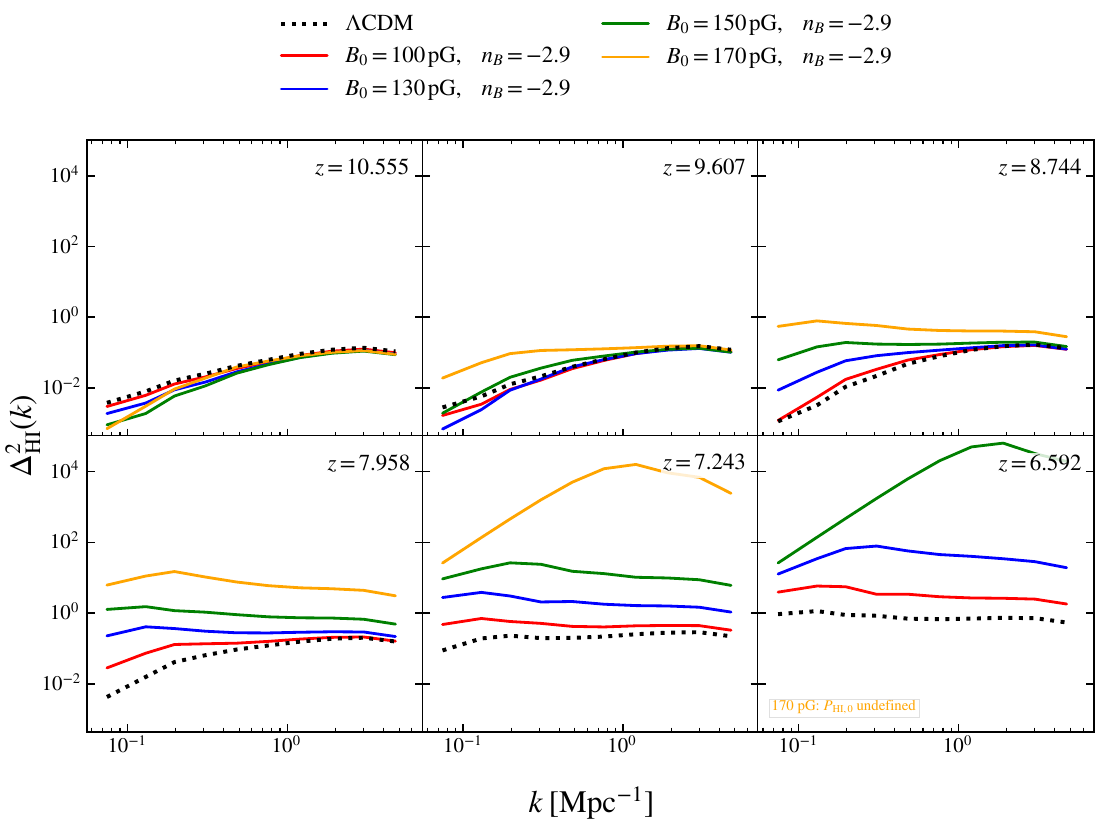}
\caption{Spherically averaged redshift-space H\,I density power spectrum for $\Lambda$CDM and the $100$, $130$, $150$, and $170\,\mathrm{pG}$ models at the indicated redshifts. The PMF models have $n_B=-2.9$. Each H\,I density contrast is defined relative to the mean H\,I density of its own model and snapshot. For the $170\,\mathrm{pG}$ model at $z=6.592$, $\overline{x}_{\rm HI}^{\,V} = 0$, so the power spectrum vanishes, as indicated by the annotation.}
\label{fig:appendix-hi-power}
\end{figure}

Figure~\ref{fig:appendix-hi-power} shows the spherically averaged redshift-space H\,I density power spectrum for the $\Lambda$CDM reference and the four PMF models. For each model and snapshot with non-zero mean H\,I density, the density contrast is defined relative to the mean of that field. We plot the dimensionless power spectrum of this fractional H\,I density contrast, $\Delta_{\rm HI}^{2}(k)=k^{3}P_{\delta_{\rm HI}}(k)/(2\pi^{2})$, where $P_{\delta_{\rm HI}}(k)$ is the spherically averaged H\,I power spectrum.

At $z=10.555$ the PMF spectrum is mostly below or near the reference for the range shown. At $z=9.607$ the $170\,\mathrm{pG}$ model has a larger fractional power on the largest scales shown, while the spectrums are closer on the small scales. At $z=8.744$ and $7.958$, the stronger-field models show a greater separation from the reference, particularly on large scales. At $z=7.243$, the $170\,\mathrm{pG}$ spectrum reaches values of the order $10^{4}$ over part of the displayed range. Similarly, at $z=6.592$, the $150\,\mathrm{pG}$ spectrum reaches values of order $10^{4}$ or larger.

These large fractional powers occur when very little neutral hydrogen remains. In particular, the mass-averaged neutral fraction is approximately $1.82\times10^{-5}$ for the $170\,\mathrm{pG}$ model at $z=7.243$ and $3.48\times10^{-6}$ for the $150\,\mathrm{pG}$ model at $z=6.592$. Normalization by the small mean H\,I density can produce large fractional fluctuations, even when the absolute brightness-temperature power is weak. The distribution and numerical sampling of the remaining neutral gas can also affect the measured late-time spectrum.

For the brightness-temperature prescription in section~\ref{sec:21cm}, the corresponding power at $k>0$ is obtained by multiplying the fractional H\,I power by
$T_0^2(z)[\overline{x}_{\rm HI}^{\,M}(z)]^2$. Consequently, a large $\Delta_{\rm HI}^{2}$ does not by itself imply a large brightness-temperature power.

At $z=6.592$, the stored H\,I volume for the $170\,\mathrm{pG}$ model is zero. Its absolute brightness power therefore vanishes, but its fractional H\,I density contrast and normalized power are undefined because the mean H\,I density is zero.


\section{Supplementary Exploration at $n_B=-2.995$}
\label{app:pmf-large-amplitude}
This appendix shows the supplementary input matter power spectrum and optical-depth results for the nearly scale-invariant sequence $n_B=-2.995$. This shows us how the model response changes over an amplitude range which is wider than that in our main-text examples. It contains exploratory amplitudes and is not restricted to models that satisfy existing observational bounds.

\subsection{Scale Dependence of the Input Matter Power Spectrum}
\label{app:pmf-spectra}
\begin{figure}
\centering
\includegraphics[width=\textwidth]
{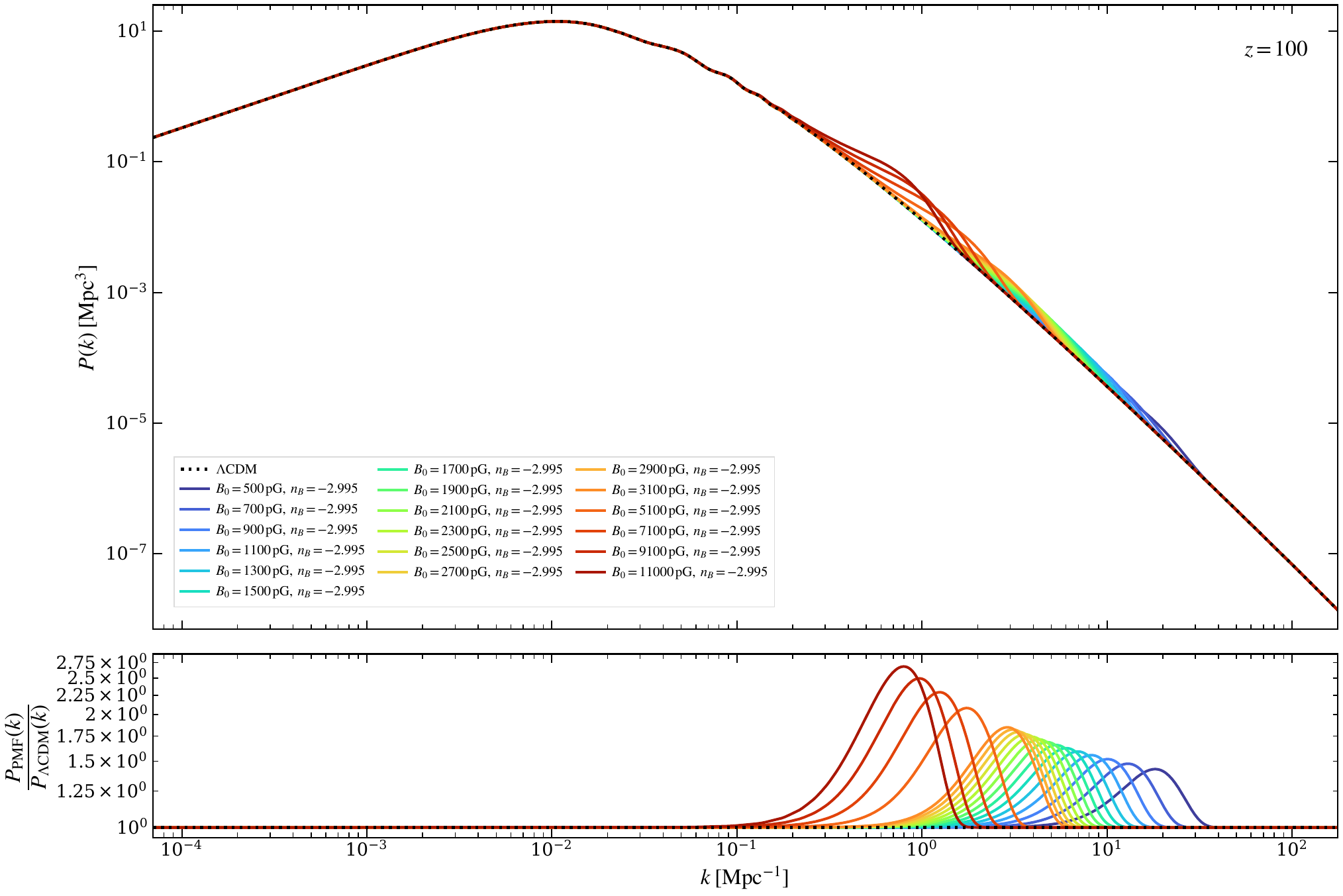}
\caption{Input linear matter power spectrum at $z=100$ for the $\Lambda$CDM reference and 18 PMF models spanning $B_0=500$--$11000\,\mathrm{pG}$ at fixed $n_B=-2.995$. The maximum relative enhancement moves towards smaller wavenumbers as the magnetic amplitude increases.}
\label{fig:app-pmf-input}
\end{figure}

Figure~\ref{fig:app-pmf-input} shows the input linear matter spectrum at $z=100$ for 18 PMF models with $B_0=500$-$11000\,\mathrm{pG}$ ($0.5$-$11\,\mathrm{nG}$), at fixed $n_B=-2.995$. Each spectrum is compared with the same $\Lambda$CDM reference. Increasing $B_0$ changes both the magnitude and the characteristic scale of the enhancement. In particular, the maximum of the ratio $P_m^{\rm PMF}(k)/P_m^{\Lambda{\rm CDM}}(k)$ moves towards smaller wavenumbers, corresponding to larger comoving scales. This maximal ratio is to be distinguished from a peak in the total matter spectrum or in the PMF-induced contribution only. The spectrum therefore do not differ merely by a uniform amplitude rescaling.

The implications of these changes for halo formation depend on the variance filtered on the Lagrangian scale corresponding to each halo mass. The top-hat window weights a wide range of wavenumbers. Consequently, a shift of the relative enhancement towards smaller $k$ does not by itself imply that the additional power ceases to contribute to the formation of low-mass halos. The analytical mass-function response depends on both $\sigma(M,z)$ and its mass derivative, while the reionization calculation depends on the resulting catalogue sources and their assigned ionization efficiencies.

These spectrums provide a possible starting point for interpreting the optical-depth response below, but do not independently establish its cause. The connection to amplitude-dependent magnetic scales and its implications for parameter selection are discussed in Section~\ref{sec:discussion-conclusions}.
%


\subsection{Optical-Depth Response}
\label{app:pmf-tau}
\begin{figure}
\centering
\includegraphics[width=\textwidth]
{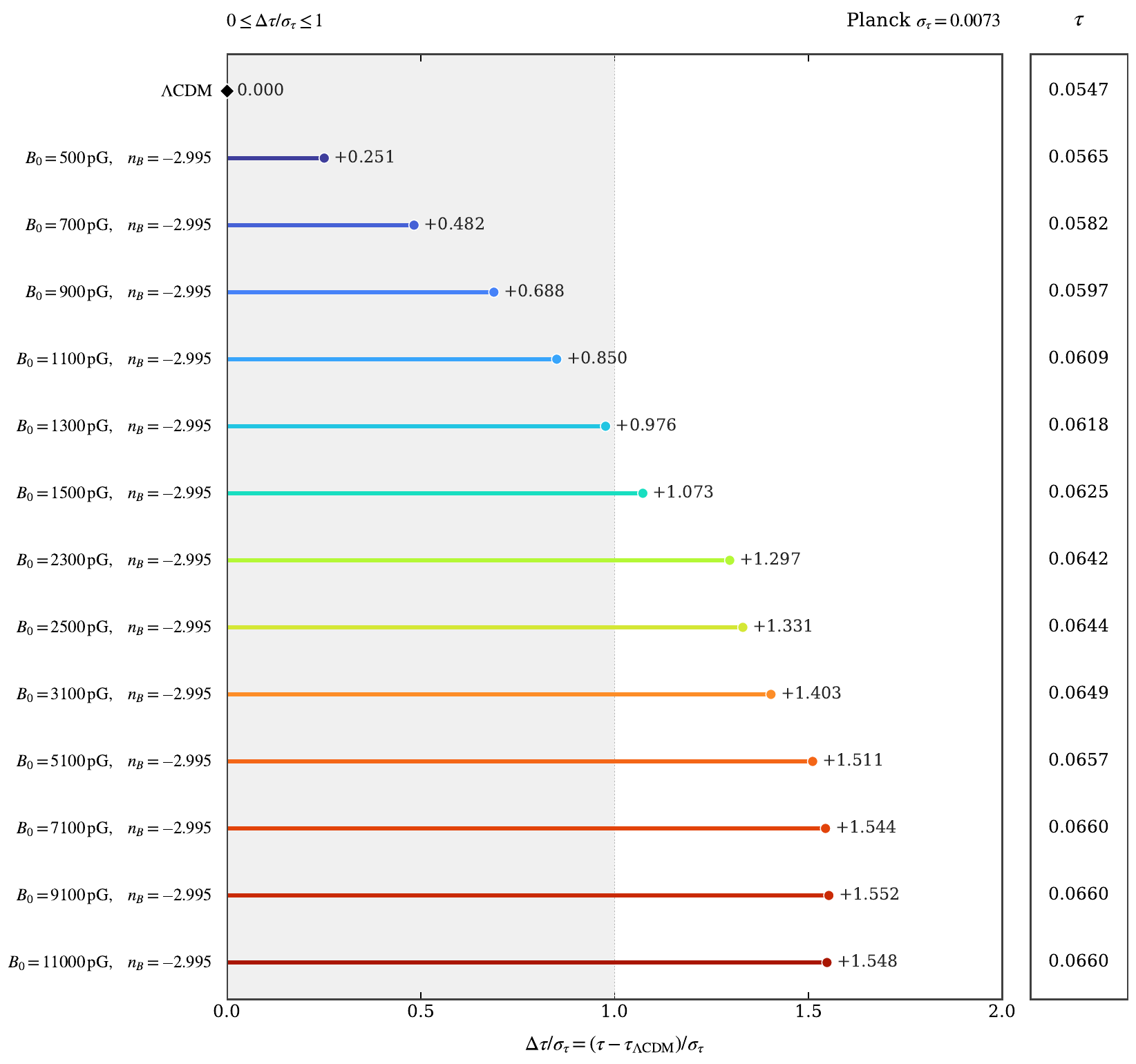}
\caption{Thomson optical-depth comparison for $\Lambda$CDM and 13 PMF models at fixed $n_B=-2.995$. The plotted coordinate is the simulation-relative offset $S_\tau$, using $\sigma_\tau=0.0073$; the optical-depth values are also listed in the figure. The largest sampled offset is $S_\tau\simeq1.5525$ at $B_0=9100\,\mathrm{pG}$.}
\label{fig:app-pmf-tau}
\end{figure}

Figure~\ref{fig:app-pmf-tau} presents optical-depth results for 13 PMF models at fixed $n_B=-2.995$ with varying $B_0$, together with the simulated $\Lambda$CDM reference. $S_\tau$ serves only as a common scale for comparing simulations. Since the subtraction uses the simulated reference, it is not an observational exclusion significance.

The optical depth increases from $\tau_{\rm e}\simeq 0.0565$ at $B_0=500\,\mathrm{pG}$ to the largest sampled value, $\tau_{\rm e} \simeq0.0660 $, at $B_0=9100\,\mathrm{pG}$. At $11000\,\mathrm{pG}$, it is slightly smaller. Using the rounded optical depths yields a maximum sampled offset of $S_\tau\simeq 1.5525$. For comparison, the offsets at $5100$ and $11000\,\mathrm{pG}$ are approximately $1.5107$ and $1.5477$, respectively. The response thus becomes progressively weaker over the sampled high-amplitude range. The small decrease between the final two amplitudes does not establish a significant physical turnover without an assessment of numerical and modeling uncertainties. The calculation uses the same optical-depth prescription as the main analysis.
\FloatBarrier

These simulations trace the response to increasing magnetic amplitude at fixed $n_B$; they do not constrain the spectral index or establish that the tested amplitudes are consistent with independent observational constraints. In particular, a small $S_\tau$ indicates a small optical-depth offset from the simulated $\Lambda$CDM reference, rather than in agreement with the full set of observational constraints. We discuss the published limits and the motivation for exploring a second spectral index in section~\ref{sec:discussion-conclusions}.



\bibliographystyle{JHEP}
\bibliography{biblio}

\end{document}